\documentclass[%
 reprint,
superscriptaddress,
 amsmath,amssymb,
 aps,
pra,
]{revtex4-2}

\DeclareMathOperator{\sinc}{sinc}

\usepackage{graphicx}
\usepackage{bm}
\usepackage{hyperref}
\usepackage{esint}
\usepackage{commath}
\usepackage{wrapfig}
\usepackage{hyperref}
\hypersetup{
    colorlinks=true,
    linkcolor=blue,
    filecolor=blue,
    urlcolor=blue,
    citecolor=blue,
    }

\begin{document}
\title{Making entangled photons indistinguishable by a time lens}

\author{Shivang Srivastava}\email{shivang.srivastava@cnrs.fr}
\affiliation{Univ. Lille, CNRS, UMR 8523 - PhLAM - Physique des Lasers Atomes et Mol\'{e}cules, F-59000 Lille, France}
\author{Dmitri B. Horoshko}\email{horoshko@ifanbel.bas-net.by}
\affiliation{Univ. Lille, CNRS, UMR 8523 - PhLAM - Physique des Lasers Atomes et Mol\'{e}cules, F-59000 Lille, France}
\affiliation{B. I. Stepanov Institute of Physics, NASB, Nezavisimosti Ave 68, Minsk 220072 Belarus}
\author{Mikhail I. Kolobov}\email{mikhail.kolobov@univ-lille.fr}
\affiliation{Univ. Lille, CNRS, UMR 8523 - PhLAM - Physique des Lasers Atomes et Mol\'{e}cules, F-59000 Lille, France}
\date{\today}

\begin{abstract}
We propose an application of quantum temporal imaging to restoring the indistinguishability of the signal and the idler photons produced in type-II spontaneous parametric down-conversion with a pulsed broadband pump. It is known that in this case, the signal and the idler photons have different spectral and temporal properties. This effect deteriorates their indistinguishability and the visibility of the Hong-Ou-Mandel interference, respectively. We demonstrate that inserting a time lens in one arm of the interferometer and choosing properly its magnification factor restores perfect indistinguishability of the signal and the idler photons and provides 100\% visibility of the Hong-Ou-Mandel interference in the limit of high focal group delay dispersion of the time lens.
\end{abstract}

\maketitle

\section{Introduction}

Classical temporal imaging is a technique of manipulation of ultrafast temporal optical waveforms similar to the manipulation of spatial wavefronts in conventional spatial imaging \cite{Copmany11, Salem:13}. It is based upon the so-called space-time duality or the mathematical equivalence of the equations describing the propagation of the temporal pulses in dispersive media and the diffraction of the spatial wavefronts in free space. Temporal imaging was first discovered in  purely electrical systems, then extended to optics \cite{Akhmanov69}, and later converted into all-optical technologies using the development of non-linear optics and ultrashort-pulse lasers \cite{Kolner88,Kolner89,Kolner94}. In the past two decades, classical temporal imaging has become a very popular tool for manipulating ultrafast temporal waveforms with numerous applications such as temporal stretching of ultrafast waveforms and compression of slow waveforms to sub-picosecond time scales, temporal microscopes, time reversal, and optical phase conjugation.

One of the key elements in classical temporal imaging is a time lens which introduces a quadratic time phase modulation into an input waveform, similar to the quadratic phase factor in the transverse spatial dimension, introduced by a conventional lens. Nowadays, optical time lenses are based on electro-optic phase modulation (EOPM) \cite{Giordmaine68,Grischkowsky74,Kolner88}, cross-phase modulation \cite{Agrawal89,Mouradian00}, sum-frequency generation (SFG) \cite{Bennett94, Bennett99,Bennett00a, Bennett00b,Hernandez13}, or four-wave mixing (FWM)\cite{Foster08, Foster09, Okawachi09, Kuzucu09}. A temporal magnification factor of the order of 100 times has been experimentally realized.

Quantum temporal imaging is a recent topic of research which brings the ideas from the spatial quantum imaging \cite{Lugiato02,Shih07,Kolobov-Book} into the temporal domain. Quantum temporal imaging searches for such manipulations of non-classical temporal waveforms which preserve their nonclassical properties such as squeezing, entanglement, or nonclassical photon statistics. Some works have already been published on this subject. Schemes for optical waveform conversion preserving the nonclassical properties such as entanglement \cite{Kielpinski11} and for aberration-corrected quantum temporal imaging of a coherent state \cite{Zhu13} have been proposed. Spectral bandwidth compression of light at a single-photon level has been experimentally demonstrated by SFG \cite{Lavoie13} and EOPM \cite{Karpinski17,Sosnicki18,Sosnicki20}. Temporal imaging in the single-photon regime has been demonstrated with an atomic-cloud-based quantum memory \cite{Mazelanik20,Mazelanik22}. Quantum temporal imaging has been demonstrated for one photon of an entangled photon pair by SFG \cite{Donohue16} and for both photons by EOPM \cite{Mittal17}. Quantum temporal imaging of broadband squeezed light was studied for SFG  \cite{Patera15,Patera17,Patera18}, and FWM-based \cite{Shi17} lenses. In Ref.~\cite{Shi20}, it was demonstrated that a time lens can preserve nonclassical effects such as antibunching and sub-Poissonian statistics of photons. In Ref.~\cite{Joshi22} a temporal magnification of two weak coherent pulses with a picosecond-scale delay by FWM was reported.

In this paper we consider the application of quantum temporal imaging to the light produced in frequency-degenerate type-II spontaneous parametric down-conversion (SPDC) pumped by a pulsed broadband source. This type of SPDC was considered in Refs.~\cite{Grice97,Keller97,Grice98}. In particular, it was demonstrated that the temporal and spectral properties of the signal and the idler down-converted photons can be significantly different in the case of a pulsed pump. This difference affects the indistinguishability of these photons and, consequently, deteriorates the visibility of the Hong-Ou-Mandel interference \cite{Hong87}. One could try to increase the indistinguishability by inserting a spectral filter in one of the arms of the interferometer. However, this filtering will result in additional losses of the SPDC light. We suggest a non-destructive way of increasing the indistinguishability by using a time lens in one of the arms instead of a filter. We demonstrate that by choosing properly the magnification factor of the temporal imaging system, one can achieve unit visibility in the Hong-Ou-Mandel interferometry, thus, reestablishing the perfect indistinguishability of the photons. We consider the case of two spectrally entangled photons only. Such photons become indistinguishable from one another after the time lens, however, they remain entangled and therefore distinguishable from other photons having the same spectral and temporal shapes. Note, that applications of single photons in boson samplers \cite{Spring13,Shchesnovich20} and quantum networks \cite{Ansari18,Ashby20,Karpinski21} require that the photons are indistinguishable and disentangled.

We employ a more realistic treatment of the time lens with respect to a usual approach found in the literature. Precisely, instead of considering the field transformation by the time lens using a local time in a reference frame traveling with the group velocity, we use the absolute time in the laboratory reference frame. This description allows us to evaluate correctly different time delays in the interferometric scheme used for observation of the Hong-Ou-Mandel effect. We clarify the role of the synchronization condition between the pulsed photon-pair source and the time lens and investigate the sensitivity of the visibility of the Hong-Ou-Mandel interference with respect to the precision of this synchronization.

The paper is organized as follows. In Sec. \ref{sec:SOI} we describe the SPDC process with a pulsed broadband pump and provide the corresponding Heisenberg equations. We evaluate the spectra and the average intensities of the signal and the idler waves. We also describe the time lens and different time delays in the interferometric scheme. In Sec. \ref{sec:CCR} we evaluate explicitly the coincidence counting rate in the Hong-Ou-Mandel interferometer and investigate the visibility of the interference pattern as a function of various physical parameters of the problem. In Sec. \ref{sec:D} we summarize our results and give a short outlook for future work.

\section{Second-order interference of photon pairs \label{sec:SOI}}

\begin{figure*}[ht]
\centering
\includegraphics[width=\linewidth]{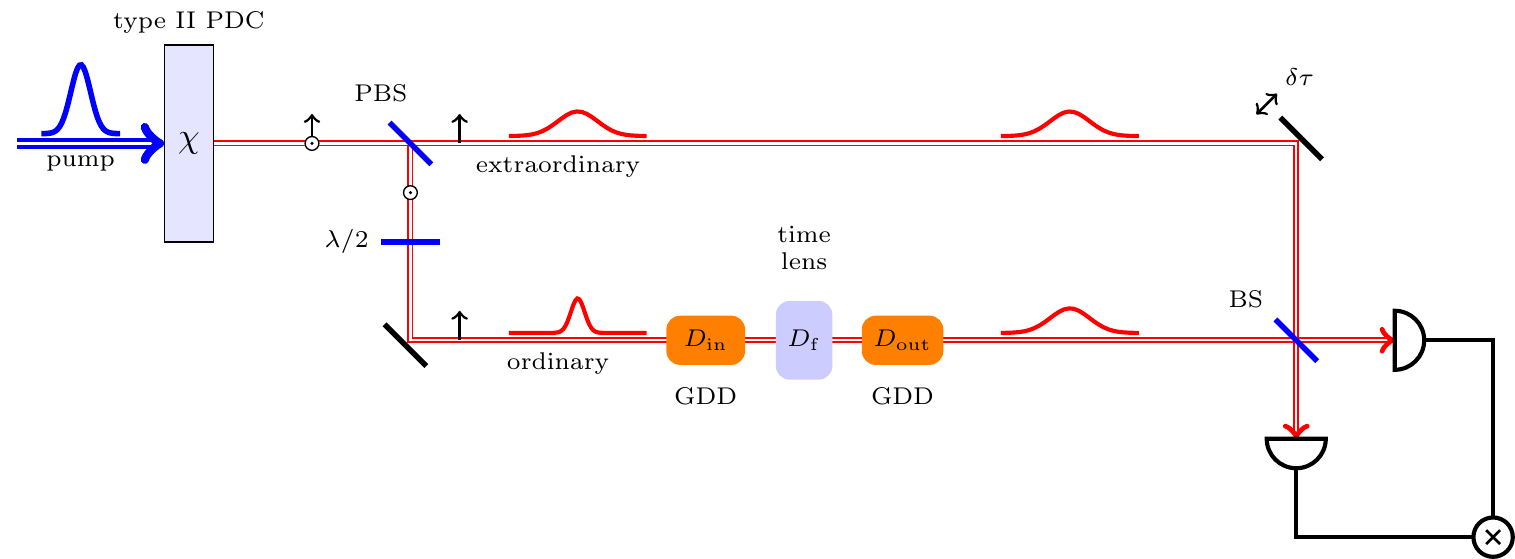}
\caption{Scheme for generation of photon pairs and observation of their second-order interference. A pump pulse impinges on a nonlinear crystal cut for type-II PDC. Two subharmonic pulses appear as ordinary and extraordinary waves in the crystal. The pump is removed, while two subharmonic pulses are separated by a polarized beam splitter (PBS). Polarization of the ordinary pulse is rotated by a half waveplate. Subsequently, the ordinary pulse passes through a temporal imaging system, composed of an input dispersive medium, a time lens, and an output dispersive medium. The delay of the extraordinary pulse $\delta\tau$ is controlled by a delay line. The pulses interfere at a beam splitter (BS) and are detected by single-photon detectors. A coincidence is registered, if both detectors fire in the same excitation cycle.  \label{fig:setup}}
\end{figure*}

The scheme for the generation of photon pairs and observation of their second-order interference is shown in Fig.~\ref{fig:setup}. It consists of several parts, which are described separately below. A quantum description of the field transformation in each optical element is done in the Heisenberg picture, instead of the Schr\"odinger picture traditionally employed for such schemes \cite{Hong87,Grice97,Keller97,Giovannetti02,Kuzucu05,Shimizu09}, because the time-lens formalism is developed in the Heisenberg picture \cite{Patera17,Patera18}. In addition, the Heisenberg picture formalism provides a natural extension to the high-gain regime of parametric downconversion (PDC), where multiple photon pairs are created at once.

\subsection{Parametric downconversion \label{sec:PDC}}

The model for the description of single-pass pulsed PDC in a $\chi^{(2)}$ nonlinear crystal in the Heisenberg picture is developed in Refs.~\cite{Caves87,Gatti03,Horoshko12,Gatti12,Horoshko19,LaVolpe21} and here we adopt its single-spatial-dimension version. We consider a crystal slab of length $L$, infinite in the transverse directions, cut for type-II collinear phase matching. We take the $x$ axis as the pump-beam propagation direction and choose the $z$ axis so that the optical axis (axes) of the crystal lies (lie) in the $xz$ plane.

The pump is a plain wave polarized in either the $y$ or $z$ direction. In the time domain, it is a Gaussian transform-limited pulse whose maximum passes the position $x=0$ at time $t=t_0$. Its central frequency is denoted by $\omega_p$. The pump is treated as an undepleted deterministic wave and is described by a $c$-number function of space and time coordinates. The positive frequency part of the pump field (in photon flux units) can be written as
\begin{equation}\label{FourierPump}
E^{(+)}_p(t,x) = \int \alpha(\Omega)e^{ik_{p}(\Omega)x -i(\omega_p+\Omega)t} \frac{d\Omega}{2\pi},	
\end{equation}
where $\Omega$ represents the frequency detuning from the central frequency and the integration limits can be extended to infinity. The pump spectral amplitude $\alpha(\Omega)$ is nonzero in a limited band only and does not depend on $x$, since the pump is undepleted. All variations of the pump wave in the longitudinal direction are determined by the wave-vector $k_p(\Omega)=n_p(\omega_p+\Omega)(\omega_p+\Omega)/c$, where $n_p(\omega)$ is the refractive index corresponding to the polarization of the pump and $c$ is the speed of light in vacuum.

We assume that the pump pulse is a transform-limited Gaussian pulse of full width at half maximum (FWHM) $\tau_p$, that is
\begin{equation}\label{Pump}
E^{(+)}_p(t,0) = E_0e^{-(t-t_0)^2/4\sigma_t^2-i\omega_pt},	
\end{equation}
where $\tau_p=2\sqrt{2\ln 2}\sigma_t$ and $E_0$ is the peak amplitude. From this equation and Eq. (\ref{FourierPump}) we find
\begin{equation}\label{alpha}
    \alpha(\Omega) =E_{0} \frac{\sqrt{\pi}}{\Omega_{p}}e^{-\Omega^{2}/4\Omega_{p}^{2}+i\Omega t_{0}},
\end{equation}
where $\Omega_{p}=1/2\sigma_{t}$.

As a result of the nonlinear transformation of the pump field in the crystal, a subharmonic field emerges. In the case of frequency-degenerate type-II phase matching, considered here, there are two subharmonic waves with the central frequency  $\omega_0=\omega_p/2$, polarized in the $y$ (ordinary wave) and $z$ (extraordinary wave) directions. These waves are treated in the framework of quantum theory and are described by Heisenberg operators. The positive-frequency part of the Heisenberg field operator (in photon flux units) can be written in the form of a Fourier integral
\begin{equation}\label{FourierSignal}
\hat E_\mu^{(+)}(t,x) = \int  \epsilon_\mu(\Omega,x)e^{ik_\mu(\Omega)x-i(\omega_0+\Omega)t} \frac{d\Omega}{2\pi},		
\end{equation}
where $\epsilon_\mu(\Omega,x)$ is the annihilation operator of a photon at position $x$ with the frequency $\omega_0+\Omega$ and polarization along the $y$ axis for the ordinary ($\mu=o$) wave or along the $z$ axis for the extraordinary ($\mu=e$) one, and  $k_\mu(\Omega)=n_\mu(\omega_0+\Omega)(\omega_0+\Omega)/c$ is the wave vector, with $n_\mu(\omega)$ the refractive index corresponding to the polarization $\mu$. The evolution of this operator along the crystal is described by the spatial Heisenberg equation \cite{Shen67}
\begin{equation}\label{evolution}
    \frac{d}{dx}\epsilon_\mu(\Omega,x) = \frac{i}\hbar\left[\epsilon_\mu(\Omega,x),G(x)\right],
\end{equation}
where the spatial Hamiltonian $G(x)$ is given by the momentum transferred through the plane $x$ \cite{Horoshko22} and equals
\begin{equation}\label{G}
    G(x) = \chi\int\limits_{-\infty}^{+\infty} E^{(-)}_p(t,x)\hat E^{(+)}_o(t,x)\hat E^{(+)}_e(t,x)dt + \mathrm{H.c.},
\end{equation}
where $\chi$ is the nonlinear coupling constant and $E^{(-)}_p(t,x)=E^{(+)*}_p(t,x)$ is the negative-frequency part of the field. Substituting Eqs.~(\ref{FourierPump}), (\ref{FourierSignal}), and (\ref{G}) into Eq.~(\ref{evolution}), performing the integration, and using the canonical equal-space commutation relations \cite{Huttner90,Horoshko22}
\begin{equation}\label{commutator}
\left[\epsilon_\mu(\Omega,x),\epsilon_\nu^\dagger(\Omega',x)\right]
= 2\pi\delta_{\mu\nu}\delta(\Omega-\Omega'),
\end{equation}
we obtain the spatial evolution equations
\begin{equation}\label{evolution2}
\begin{split}
    \frac{d\epsilon_o(\Omega,x)}{dx} &=\kappa\int \alpha(\Omega+\Omega')\epsilon^{\dagger}_e(\Omega',x)e^{i\Delta(\Omega,\Omega')x}d\Omega',\\
    \frac{d\epsilon_e(\Omega,x)}{dx} &=\kappa\int \alpha(\Omega+\Omega')\epsilon^{\dagger}_o(\Omega',x)e^{i\Delta(\Omega',\Omega)x}d\Omega',
\end{split}
\end{equation}
where $\kappa=i\chi/2\pi\hbar$ is the new coupling constant and $\Delta(\Omega,\Omega')=k_p(\Omega+\Omega')-k_o(\Omega)-k_e(\Omega')$ is the phase mismatch for the three intracting waves.

In the low-gain regime, where the pump amplitude is sufficiently small, we can solve Eq.~(\ref{evolution2}) perturbatively, by substituting $\epsilon^{\dagger}_\mu(\Omega',x)\to\epsilon^{\dagger}_\mu(\Omega',0)$ under the integral. In this way, we obtain for the fields at the crystal output
\begin{equation}\label{CrystalSolution}
\begin{split}
    \epsilon_o(\Omega,L)&=\epsilon_o(\Omega,0)\\ &\quad+\kappa L\int \alpha(\Omega+\Omega')
  \epsilon^{\dagger}_e(\Omega',0) \Phi(\Omega,\Omega')d\Omega',\\
   \epsilon_e(\Omega,L)&=\epsilon_e(\Omega,0)\\ &\quad+\kappa L\int \alpha(\Omega+\Omega')
  \epsilon^{\dagger}_o(\Omega',0) \Phi(\Omega',\Omega)d\Omega',
\end{split}
\end{equation}
where $L$ is the crystal length and
\begin{equation}\label{Phidef}
\Phi(\Omega,\Omega')=e^{i\Delta(\Omega,\Omega')L/2}\sinc[\Delta(\Omega,\Omega')L/2]
\end{equation}
is the phase-matching function.

Substituting these solutions into Eq. (\ref{FourierSignal}), we obtain the field transformation from the crystal input face to its output one. To write this transformation in a compact form, we define the envelopes of the ordinary wave at the crystal input and output faces as $A_0(t)=\hat E_o^{(+)}(t,0) e^{i\omega_0t}$ and $A_1(t)=\hat E_o^{(+)}(t,L)e^{i(\omega_0t-k^0_oL)}$ respectively, and those of the extraordinary wave at the same positions as $B_0(t)=\hat E_e^{(+)}(t,0)e^{i\omega_0t}$ and $B_1(t)=\hat E_e^{(+)}(t,L)e^{i(\omega_0t-k_e^0L)}$ respectively. Here $k_\mu^0=k_\mu(0)$.  Using this notation, the field transformation in the crystal has the form of an integral Bogoliubov transformation
\begin{eqnarray}\label{BogoliubovA}
A_1(t) &=& \int U_A(t,t')A_0(t')dt' + \int V_A(t,t')B_0^\dagger(t')dt',\\\label{BogoliubovB}
B_1(t) &=& \int U_B(t,t')B_0(t')dt' + \int V_B(t,t')A_0^\dagger(t')dt',
\end{eqnarray}
where the Bogoliubov kernels are
\begin{eqnarray}\label{UA}
U_A(t,t') &=& \int e^{i[k_o(\Omega)-k_o^0]L+i\Omega(t'-t)}\frac{d\Omega}{2\pi},\\\label{VA}
V_A(t,t') &=&\int e^{i[k_o(\Omega)-k_o^0]L-i(\Omega't'+\Omega t)}J(\Omega,\Omega')\frac{d\Omega d\Omega'}{2\pi} ,\\\label{UB}
U_B(t,t') &=& \int e^{i[k_e(\Omega)-k_e^0]L+i\Omega(t'-t)}\frac{d\Omega}{2\pi},\\\label{VB}
V_B(t,t') &=& \int e^{i[k_e(\Omega')-k_e^0]L-i(\Omega't+\Omega t')} J(\Omega,\Omega')\frac{d\Omega d\Omega'}{2\pi}.
\end{eqnarray}
and
\begin{equation}\label{J}
J(\Omega,\Omega')=
\kappa L\alpha(\Omega+\Omega')\Phi(\Omega,\Omega')
\end{equation}
is the joint spectral amplitude (JSA) of the two generated photons, or the biphoton \cite{KlyshkoBook}.

\subsection{Spectral and temporal shapes of the photons \label{sec:shapes}}

The spectra of the ordinary and extraordinary waves are $S_\mu(\Omega) = \langle\epsilon_\mu^\dagger(\Omega,L)\epsilon_\mu(\Omega,L)\rangle$, where $\mu=o,e$. Substituting the solutions (\ref{CrystalSolution}), applying the commutation relation (\ref{commutator}), and setting to zero all normally ordered averages at $x=0$, we obtain
\begin{equation}\label{So}
S_o(\Omega) =  2\pi\int \left|J(\Omega,\Omega')\right|^2 d\Omega',
\end{equation}
and a similar expression for $S_e(\Omega)$ with $J(\Omega,\Omega')$ replaced by $J(\Omega',\Omega)$.

To find these spectra analytically, we make two approximations. The first one is the approximation of linear dispersion in the crystal. This means that we consider only linear terms in the dispersion law in the crystal, which is justified for a not-too-long crystal \cite{Grice97,Grice01}. Thus we write  $k_\mu(\Omega) \approx k_{\mu}^0 + k_\mu'\Omega$, where $k_\mu'=(d k_\mu/d \Omega)_{\Omega=0}$ is the inverse group velocity of the wave $\mu=p,o,e$ in the nonlinear crystal. In this way we obtain
%
$  \Delta(\Omega,\Omega')L/2 \approx \tau_o\Omega+\tau_e\Omega'$,
%
where $\tau_o=(k_p'-k_o')L/2$ and $\tau_e=(k_p'-k_e')L/2$ are relative group delays of the ordinary and extraodinary photons with respect to the pump at half crystal length and we have also assumed a perfect phase matching at degeneracy, $k_p^0-k_o^0-k_e^0=0$.

The second approximation is the replacement of the $\sinc(x)$ function by a Gaussian function having the same width at half maximum $e^{-x^2/2\sigma_s^2}$, where $\sigma_s=1.61$ \cite{Grice01,LaVolpe21}. Applying both approximations, we write
%
\begin{equation}\label{approxPhi}
    \Phi(\Omega,\Omega')\approx\exp\left[-\frac{(\tau_o\Omega+\tau_e\Omega')^2}{2\sigma_{s}^{2}}+i(\tau_o\Omega+\tau_e\Omega')\right].
\end{equation}

Substituting Eqs. (\ref{alpha}), (\ref{approxPhi}) and (\ref{J}) into Eq. (\ref{So}) and a similar expression for $S_e(\Omega)$ and performing integrations (see Appendix \ref{sec:appendixa}), we find
\begin{equation}\label{Smu}
S_\mu(\Omega) = \frac{\sqrt{2\pi}P_b}{\sigma_\mu} e^{-\Omega^2/2\sigma_\mu^2},
\end{equation}
where the spectral standard deviation of the ordinary photon is
\begin{equation}\label{sigmao}
\sigma_o = \frac{\sqrt{\sigma_s^2+2\tau_e^2\Omega_p^2}}{\sqrt{2}|\tau_e-\tau_o|},
\end{equation}
that of the extraordinary photon, $\sigma_e$, is obtained by a replacement $\tau_o\leftrightarrow\tau_e$, and
\begin{equation}
P_b = \int \left|J(\Omega,\Omega')\right|^2 d\Omega d\Omega' = \frac{\sqrt{2}\pi^2(\kappa LE_0)^2\sigma_s}{\Omega_p|\tau_e-\tau_o|}
\end{equation}
is the probability of biphoton generation per pump pulse.

\begin{figure}[ht]
\centering
\includegraphics[width=\linewidth]{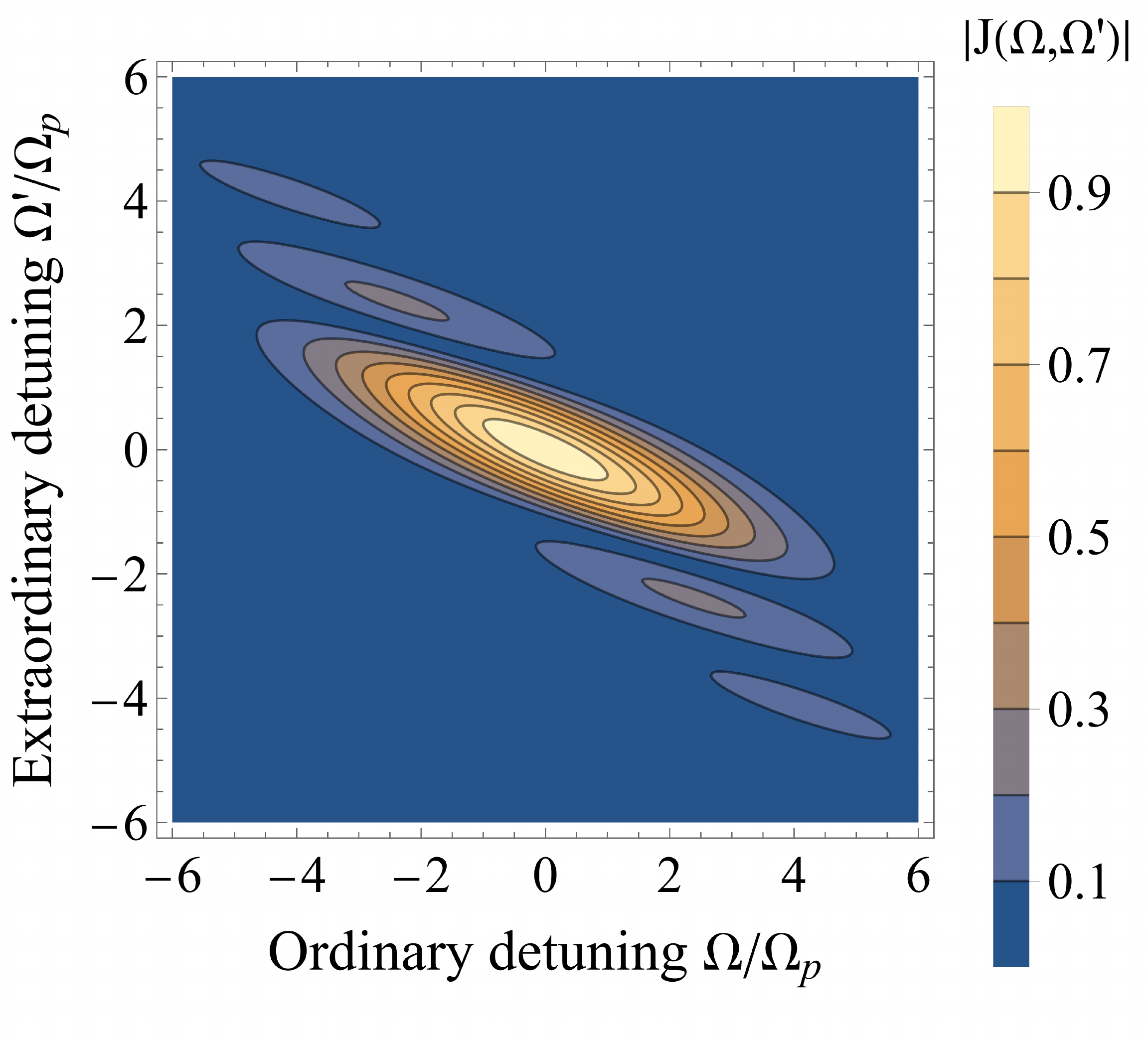}
\caption{Normalized contour plot of the JSA for two photons generated in a 2-cm-long BBO crystal pumped at 405 nm. The tilted shape of the JSA indicates the frequency-time entanglement between the photons. The bandwidth of the ordinary photon is more than two times higher than that of the extraordinary one. \label{fig:jsa}}
\end{figure}

In Fig.~\ref{fig:jsa} we show the modulus of the JSA for a $\beta$ barium borate (BBO) crystal of length $L=2$ cm, pumped at 405 nm by Gaussian pump pulses with a FWHM bandwidth $\Delta\lambda=0.2$ nm, similar to the experimental setup of Ref.~\cite{Grice98}, but with a longer crystal and longer pump pulses. The pump bandwidth corresponds to $\Omega_p=0.98$ rad/ps  or $\tau_p=1.21$ ps. Using the Sellmeier equations for the ordinary and extraordinary refractive indices of BBO \cite{Eimerl87}, we find the angle between the pump and the optical axis $\theta_p=41.42^{\circ}$ for a frequency-degenerate collinear phase matching, $\tau_o= 0.76$ ps and $\tau_e=2.68$ ps. For these conditions, we find $\sigma_o=1.48$ rad/ps and $\sigma_e=0.71$ rad/ps, which give the ratio $\sigma_o/\sigma_e=2.1$, showing the asymmetry in the spectral bandwidth of two generated photons.

The temporal shapes of the photons are given by their average intensities (in photon flux units) $I_\mu(t)=\langle\hat E_\mu^{(-)}(t,L)\hat E_\mu^{(+)}(t,L)\rangle$. For the ordinary photon, from Eqs. (\ref{BogoliubovA}), (\ref{UA}), and (\ref{VA}) we obtain
\begin{eqnarray}\nonumber
I_o(t) &=& \frac1{2\pi} \int e^{i[k_o(\Omega')-k_o(\Omega)]L+i(\Omega-\Omega')t}
d\Omega d\Omega'\\\label{Io}
&\times&\int J^*(\Omega,\Omega'')J(\Omega',\Omega'')d\Omega''.
\end{eqnarray}

With the help of the same approximations as above, we find (see Appendix \ref{sec:appendixa})
\begin{equation}\label{Imu}
I_\mu(t) = \frac{P_b}{\sqrt{2\pi}\Delta t_\mu}e^{-(t-t_0-\tau_\mu-k_\mu'L)^2/2\Delta t_\mu^2},
\end{equation}
where the temporal standard deviation of the ordinary photon is
\begin{equation}
\Delta t_o = \frac{\sqrt{\sigma_s^2+2\tau_o^2\Omega_p^2}}{2\sigma_s\Omega_p},
\end{equation}
while that of the extraordinary photon, $\Delta t_e$, is obtained by a replacement  $\tau_o\leftrightarrow\tau_e$. We note that the ratio of the temporal widths of two photons is equal to the inverse of that of their spectral widths: $\Delta t_e/\Delta t_o=\sigma_o/\sigma_e$. For the above example of a BBO crystal, we find $\Delta t_o=0.61$ ps and $\Delta t_e=1.28$ ps.

We note also that for both photons
\begin{equation}\label{integral}
\int\limits_{-\infty}^{+\infty} I_\mu(t)dt = \int\limits_{-\infty}^{+\infty} S_\mu(\Omega)\frac{d\Omega}{2\pi} = P_b,
\end{equation}
and since $I_\mu(t)$ has the meaning of the photon flux, Eq. (\ref{integral}) justifies the interpretation of $P_b$ as the probability of biphoton generation in one excitation cycle.

For a sufficiently narrowband pump, when $\tau_\mu\Omega_p\ll\sigma_s$, the temporal standard deviations of the ordinary and extraordinary photons are approximately the same and equal to $1/2\Omega_p=\sigma_t$, which simply means that the down-converted photons have a probability to appear until the pump pulse is inside the crystal. The spectral standard deviations of both photons are the same in this case and equal to the continuous-wave (CW) spectral standard deviation
\begin{equation}
\sigma_\mathrm{cw} = \frac{\sigma_s}{\sqrt{2}|\tau_e-\tau_o|}.
\end{equation}
which gives $\sigma_\mathrm{cw}=0.59$ rad/ps for our example of a BBO crystal.

\begin{figure}[ht]
\centering
\includegraphics[width=\columnwidth]{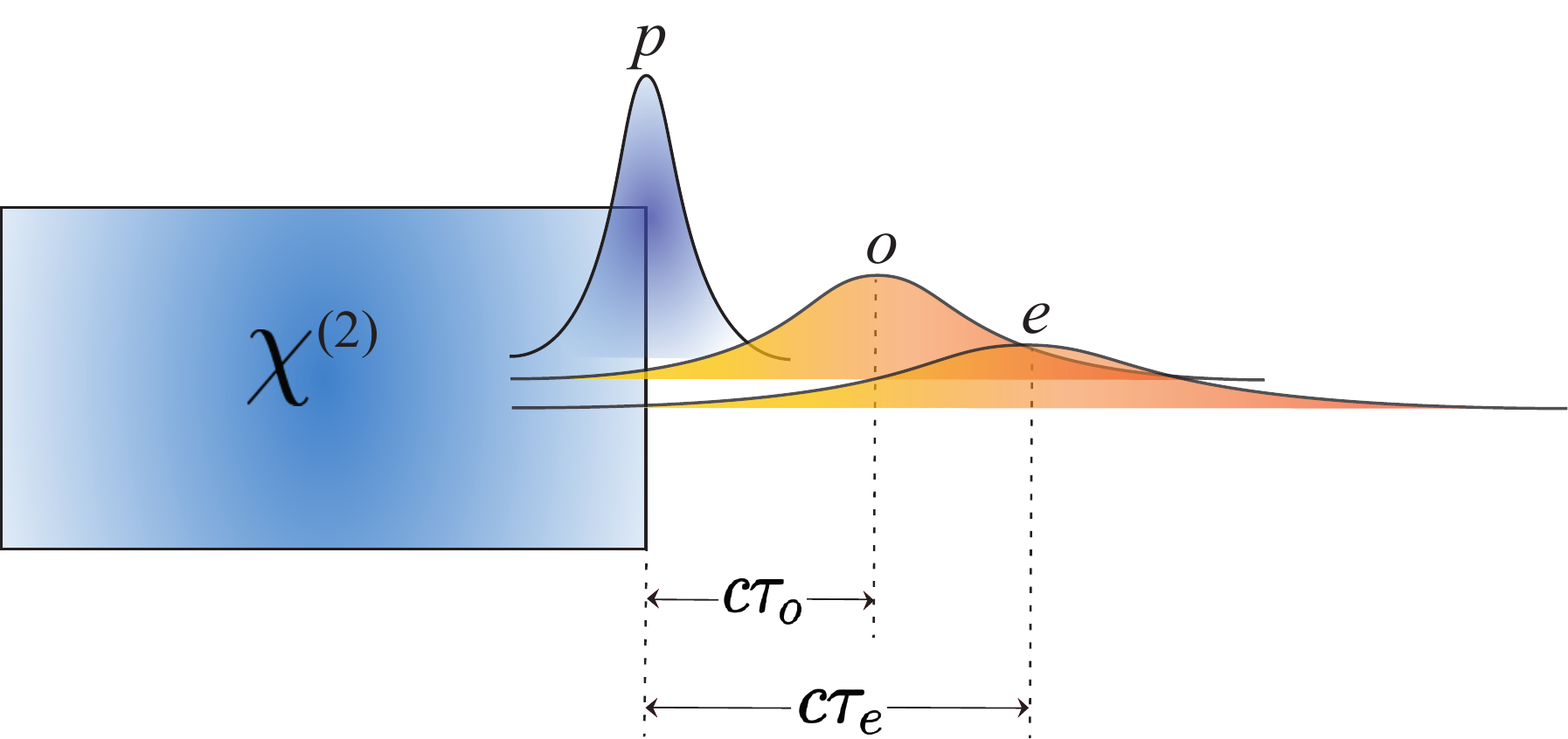}
\caption{Schematic representation of the generation of two photons, ordinary (o) and extraordinary (e), by type-II PDC from a pump pulse (p) in a crystal with quadratic nonlinear susceptibility. For a broadband pump, the generated wave packets may be much longer than the pump pulse. In a crystal with positive dispersion, the generated photons travel at a higher group velocity in comparison to the pump and advance the pump pulse at the output. \label{fig:temporal}}
\end{figure}

However, when the above condition is not satisfied, and, in addition, the group velocities of the ordinary and extraordinary waves are significantly different, the durations and delays of the two generated photons are different too, as shown schematically in Fig. \ref{fig:temporal}. The peak of the single-photon wavepacket polarized in the direction $\mu=o,e$ passes the output face of the crystal at time $t_0+\tau_\mu+k_\mu'L$, while the peak of the pump pulse passes this face at time $t_0+k_p'L$. The relative delay with respect to the pump is thus $-\tau_\mu$, it is negative in a crystal with a positive dispersion, where the group velocity decreases with frequency, which means that the downconverted photons advance the pump pulse. The relative delay between the photons can be easily compensated by an optical delay line. Still, the difference in wavepacket duration is a serious problem for observing a Hong-Ou-Mandel interference of them. The photons are partially distinguishable in this case, and, as a consequence, the Hong-Ou-Mandel interference of these photons exhibits a degraded visibility \cite{Grice97,Keller97,Grice98}. A time lens can be used for temporal stretching of the shorter wavepacket, as shown in the following sections.

\subsection{Temporal imaging system \label{sec:TI}}

The interferometer we consider includes a single-lens temporal imaging system in its arm where the ordinary wave propagates. Such a system realizes a temporal stretching or compression of a waveform similar to the action of an ordinary single-lens imaging system in space. It consists of an input dispersive medium, a time lens, and an output dispersive medium. To understand its action on the field, it is instructive to apply the quadratic approximation for the dispersion law as follows.

An optical wave with a carrier frequency $\omega_0$ passing through a transparent medium experiences dispersion characterized by the dependence of its wave vector $k(\Omega)$ on the frequency, which we decompose around the carrier frequency in powers of $\Omega=\omega-\omega_0$ and limit the Taylor series to the first three terms
\begin{equation}\label{quadratic}
     k(\Omega) \approx k_0+k_0'\Omega + \frac12k_0''\Omega^2,
\end{equation}
where $k_0=k(0)$, $k_0' = (d k/d \Omega)_{\omega_0}$ is the inverse group velocity, and $k_0'' = (d^2 k/d \Omega^2)_{\omega_0}$ is the group-velocity dispersion of the medium at the carrier frequency $\omega_0$. The effect of the first-order term of this series is the group delay by the time $k_0'l$, where $l$ is the length of the medium, as has already been discussed in the previous section. The second-order term of this series is responsible for the spreading of the initial waveform, which is characterized by the group delay dispersion (GDD) $D=k_0''l$. The input and output dispersive media of the temporal imaging system have GDDs $D_\mathrm{in}$ and $D_\mathrm{out}$ respectively. A time lens is a device realizing a quadratic-in-time phase modulation of the passing waveform. It can be realized by an EOPM (electro-optical time lens) or by a nonlinear medium performing SFG of FWM (parametric time lens). In both cases, a time lens is characterized by its focal GDD $D_\mathrm{f}$. The condition of single-lens temporal imaging reads
\begin{equation}\label{condition}
    \frac{1}{D_\mathrm{in}}+\frac{1}{D_\mathrm{out}}=\frac{1}{D_\mathrm{f}}
\end{equation}
and is similar to the thin lens equation. When this condition is satisfied, the output waveform is a magnified version of the input one with a magnification $M=-D_\mathrm{out}/D_\mathrm{in}$.

Transformation of the field envelope operator in the temporal imaging system can be found in Refs. \cite{Patera15,Patera18}, where it is written in the delayed reference frame (see also Ref. \cite{Kolner94} for its classical form). Here we rewrite this transformation in absolute time as
\begin{equation}\label{LensEq}
    A_3(t_\mathrm{out}+\Delta t_\mathrm{out}) = \frac{-1}{\sqrt{M}}
    e^{i\Delta t_\mathrm{out}^2/2MD_\mathrm{f}}A_2(t_\mathrm{in}+\Delta t_\mathrm{in}),
\end{equation}
where $A_2(t)$ and $A_3(t)$ are field envelopes at the input and output of the temporal imaging system respectively. Here $t_\mathrm{in}$ is the time corresponding to the center of the temporal field of view \cite{Shi17}, a time axis (similar to the optical axis of a spatial imaging system). This time is determined by the time of passage of the modulating signal in an EOPM or by the time of passage of the pump pulse in a parametric time lens. Correspondingly, $t_\mathrm{out}$ is the central time of the image waveform. It is related to $t_\mathrm{in}$ as $t_\mathrm{out}=t_\mathrm{in}+t_d$, where $t_d$ is the group delay time in the temporal imaging system. In addition, $\Delta t_\mathrm{out}$ and $\Delta t_\mathrm{in}$ are times relative to the centers of the object and image, respectively; they are related by a scaling $\Delta t_\mathrm{out}=M\Delta t_\mathrm{in}$.

For practical use, we rewrite Eq. (\ref{LensEq}) as
\begin{equation}\label{LensEq2}
    A_3(t) = \frac{-1}{\sqrt{M}}
    e^{i\left(t-t_\mathrm{out}\right)^2/2MD_\mathrm{f}}A_2\left(t_\mathrm{out} -t_d+\frac{t-t_\mathrm{out}}M\right),
\end{equation}
where $t_\mathrm{out}$ and $t_d$ can be considered as constants specific to a given time lens.

Equations (\ref{LensEq}) and (\ref{LensEq2}) are obtained in several approximations, the two most important being that of the Fraunhofer limit for the dispersion and that of an infinitely large temporal aperture $T_A$ of the time lens. The first approximation mentioned requires that $D_\mathrm{in}^2 \gg 1/4\sigma_o^4$, as shown in Appendix \ref{sec:appendixb}. The second one requires that the FWHM of the stretched pulse does not surpass the temporal aperture, i.e., $2\sqrt{2\ln2}\Delta t_\mathrm{str}<T_A$, where $\Delta t_\mathrm{str} = |D_\mathrm{in}|\sigma_o$ is the temporal standard deviation of the pulse stretched in the first dispersive medium, as also shown in Appendix \ref{sec:appendixb}. While the first requirement sets a lower bound for $\sigma_o$, the second one sets its upper bound. We obtain from Eq. (\ref{condition}) and the definition of magnification that $D_\mathrm{in} = D_\mathrm{f}(M-1)/M$. Therefore, we can write both bounds as
\begin{equation}\label{cond1}
  \frac{M^2}{4D_\mathrm{f}^2(M-1)^2} \ll \sigma_o^4 < \frac{T_A^4M^4}{D_\mathrm{f}^4(M-1)^4 (8\ln2)^2}.
\end{equation}
The bounds are compatible if
\begin{equation}\label{cond2}
  T_A^4 \gg  (4\ln2)^2D_\mathrm{f}^2(M-1)^2/M^2.
\end{equation}
As mentioned in the Introduction, a time lens can be realized by an EOPM or a nonlinear optical process, such as SFG or FWM. An SFG-based time lens always changes the carrier frequency of the signal field and is therefore unsuitable for the case of two photons having the same carrier frequency, considered here.

For an EOPM-based time lens, the temporal aperture is given by $T_A=1/2\pi f_\mathrm{RF}$, where $f_\mathrm{RF}$ is the operating radio frequency of the phase modulator, while the focal GDD is $D_\mathrm{f}=T_A^2/\theta_\mathrm{max}$, where $\theta_\mathrm{max}$ is the maximal achievable phase shift, limited by the damage threshold of the modulator crystal \cite{Kolner94,Copmany11,Salem:13}. Thus, Eq. (\ref{cond2}) implies $\theta_\mathrm{max}^2\gg 7.7(M-1)^2/M^2$, which is feasible for modern EOPM-based time lenses, where a value of $\theta_\mathrm{max}=25$ rad has been reported \cite{Karpinski17}, and for a magnification $|M|>1$, which we consider in the setup of Fig.~\ref{fig:setup}. Taking this value of the maximal phase shift and $f_\mathrm{RF}=40$ GHz \cite{Karpinski17,Ashby20}, we obtain $T_A=4$ ps and $D_\mathrm{f}=0.64$ ps$^2$. For a magnification $M=-2.1$, required for the case of the previous section, we obtain from Eq. (\ref{cond1}) the condition $\sigma_o^4\gg 0.28$ rad$^4$/ps$^4$ and $\sigma_o< 1.8$ rad/ps. Both these conditions are satisfied by the BBO source considered in the previous section. We note that a Fresnel time lens based on an EOPM with a nonsinusoidal driving current provides a higher temporal aperture for fixed  $\theta_\mathrm{max}$ and $D_\mathrm{f}$ \cite{Sosnicki18,Sosnicki20}.

Frequency-degenerate FWM is technically more complicated but provides a wider range of possible focal GDDs and temporal apertures. In a FWM-based time lens, a pump pulse of FWHM duration $\tau_0$ passes through a dispersive medium with GDD $D_\mathrm{p}$ and acquires a FWHM duration $T_A=4\ln2 |D_\mathrm{p}|/\tau_0$, which constitutes the temporal aperture. The focal GDD is given by $D_\mathrm{f}=-D_\mathrm{p}/2$ \cite{Salem:13}. Taking realistic values $\tau_0=0.1$ ps, $D_\mathrm{p}=-44$ ps$^2$ \cite{Kuzucu09}, and $M=-2.1$, we obtain from Eq. (\ref{cond1}) the condition $\sigma_o^4\gg 0.0002$ rad$^4$/ps$^4$ and $\sigma_o< 16$ rad/ps. Both these conditions are satisfied with very wide margins by the BBO source considered in the previous section.

\subsection{Coincidence detection}

We define the position of the output mirror of the interferometer as $x=x_4$ and assume that the detectors are placed just after it at the same position. The field operators at detectors D1 and D2 are given by
\begin{equation}\label{mirror}
\begin{split}
 &\hat{E}^{(+)}_{1}(t)=\frac{1}{\sqrt{2}}[\hat{E}^{(+)}_o(t,x_4)+\hat{E}^{(+)}_e(t-t_z,x_1)],\\
   &\hat{E}^{(+)}_{2}(t)=\frac{1}{\sqrt{2}}[\hat{E}^{(+)}_o(t,x_4)-\hat{E}^{(+)}_e(t-t_z,x_1)],
\end{split}
\end{equation}
where $t_z=t_{z0}+\delta\tau$ with $t_{z0}$ the time delay experienced by the extraordinary wave in the interferometer and $\delta\tau$ the additional delay introduced for this wave.

The probability of detecting one photon at detector D1 at
time $t_1$ and one photon at detector D2 at time $t_2$ is
\begin{eqnarray}\nonumber
    P_{12}(t_{1},t_{2};\delta\tau)&=&\langle \hat{E}^{(-)}_{1}(t_1)\hat{E}^{(-)}_{2}(t_2)\hat{E}^{(+)}_{2}(t_2)\hat{E}^{(+)}_{1}(t_1) \rangle\\\label{prob}
    &=& \left|\langle \hat{E}^{(+)}_{2}(t_2)\hat{E}^{(+)}_{1}(t_1) \rangle\right|^2 + O(\xi^4),
\end{eqnarray}
where $\xi=\kappa LE_0\ll1$ is the smallness parameter of the low-gain regime of PDC.

The average coincidence counting rate is given by \cite{Grice97}
\begin{equation}\label{Rc}
    R_{c}(\delta\tau)= \frac{1}{T}\int\int_{0}^{T}dt_{1}dt_{2}P_{12}(t_{1},t_{2};\delta\tau)
\end{equation}
where $T$ is the coincidence detection time.

\subsection{Relative delays \label{sec:Delays}}

Uniting the transformations presented above, we need to take into account relative delays between the parts of the interferometric setup shown in Fig. \ref{fig:setup}. In doing this, we need to remember that the field envelope travels in a dispersive medium at the group velocity, while the central frequency component travels at the phase velocity. In air, these two velocities practically coincide, but in a dense dispersive medium, they may be significantly different, which results in the appearance of a phase shift for the field amplitude.

In this way, we write $A_2(t)=A_1(t-t_{12})$, where $t_{12}$ is the delay between the point $x_1=L$ (PDC crystal output) and point $x_2$ (temporal imaging system input). Also, we write $\hat{E}^{(+)}_o(t,x_4)=A_3(t-t_{34})e^{-i\omega_0t+i\phi}$, where $t_{34}$ is the delay between the point $x_3$ (temporal imaging system output) and point $x_4$ (interferometer output mirror), while $\phi$ is a phase caused by the difference of the phase and group velocities in the media of the temporal imaging system. We do not write this phase explicitly, because it does not affect the second-order interference picture, as we show below.

For perfect interference, we expect that the total delay of the ordinary wave matches that of the extraordinary one. The ordinary wave emerges when the pump pulse enters the nonlinear crystal. In a crystal with positive dispersion, the subharmonic travels at a higher group velocity than the pump. This means that the ordinary wave is a longer pulse, advanced with respect to the  pump pulse (see Fig. \ref{fig:temporal}). The center of this ordinary pulse is generated when the peak of the pump pulse passes the crystal center, which happens at time $t_0+k_p'L/2$. We recall that $k_p'=1/v_g$ is the inverse of the group velocity $v_g$ of the pump. Thus, the center of the ordinary pulse leaves the crystal at time $t_0+k_p'L/2+k_o'L/2$ and arrives at the exit interferometer mirror, point $x_4$, at time
%
   $ t_{4o}=t_0+k_p'L/2+k_o'L/2+t_{12}+t_d+t_{34}$.
%
Similar reasoning gives the time of arrival of the center of the extraordinary pulse
%
$    t_{4e}=t_0+k_p'L/2+k_e'L/2+t_{z0}$.
%
A perfect interference is expected to occur at $t_{4o}=t_{4e}$, wherefrom
\begin{equation}\label{rel-interf}
   k_o'L/2+t_{12}+t_d+t_{34}=k_e'L/2 +t_{z0}.
\end{equation}

Another important relationship is the synchronization condition for the photon pair source and the modulator. The center of the temporal field of view (time axis) $t_\mathrm{in}$ is expected to coincide with the center of the ordinary pulse, arriving at the time lens:
\begin{equation}\label{rel-synchro}
  t_\mathrm{in} = t_0+k_p'L/2+k_o'L/2+t_{12} + \delta t,
\end{equation}
where $\delta t$ is a time shift, describing a possible synchronization imprecision.

Now Eqs. (\ref{BogoliubovA}), (\ref{BogoliubovB}), (\ref{LensEq2}), and (\ref{mirror}) and (\ref{prob}) represent a closed system of equations, allowing us to express the fields on the detectors through the vacuum fields at the PDC crystal input and calculate the average coincidence counting rate defined in Eq. (\ref{Rc}).

\section{Analytical calculation of the average coincidence counting rate \label{sec:CCR}}
In this section we calculate the average coincidence counting rate bringing together the relations for different parts of the interferometric setup, obtained in the preceding section. To this end, we express the fields in Eq. (\ref{prob}) via the vacuum fields $A_0(t)$ and $B_0(t)$ with the help of Eqs. (\ref{BogoliubovA}), (\ref{BogoliubovB}), (\ref{mirror}), (\ref{LensEq2}) and the relations for the field envelopes, discussed in Sec. \ref{sec:Delays}. Then we transform the obtained expression to the normal form using the commutation relations (\ref{commutator}), and leave only the $c$-number terms, since the normally ordered correlator for the vacuum field is zero. In the lowest (quadratic) order of $\xi=\kappa LE_0\ll1$, we obtain
\begin{widetext}
\begin{equation}\label{P12new}
\begin{split}
    P_{12}(t_1, t_{2};\delta\tau)
    &=\frac1{4|M|}\left| e^{i(t_{1}-t_{34}-t_\mathrm{out})^{2}/2MD_{\mathrm{f}}}\int U_{B}(t_{2}-t_{z},t')
    V_{A}(t_y+\frac{t_{1}-t_{34}-t_\mathrm{out}}{M},t')dt'\right.\\
    &\quad- \left.e^{i(t_{2}-t_{34}-t_\mathrm{out})^{2}/2MD_{\mathrm{f}}}
    \int U_{A}(t_y+\frac{t_{2}-t_{34}-t_\mathrm{out}}{M},t')V_{B}(t_{1}-t_{z},t')dt'\right|^2,
\end{split}
\end{equation}
\end{widetext}
where $t_y=t_\mathrm{out}-t_{d}-t_{12}$.

Substituting the expressions for the Bogoliubov kernels (\ref{UA})--(\ref{VB}), into Eq. (\ref{P12new}) and the obtained expression into Eq. (\ref{Rc}), extending the limits of integration over $t_1$ and $t_2$ to infinity \cite{Grice97}, and using the relations for relative delays (\ref{rel-interf}) and (\ref{rel-synchro}), we obtain
\begin{equation}\label{Rc3}
    R_c(\delta\tau) = \frac{P_b}{2T}\left[1 - p_\mathrm{int}(\delta\tau)\right],
\end{equation}
where
\begin{eqnarray}\label{pint}
p_\mathrm{int}(\delta\tau)&=&\frac{D_{\mathrm{f}}}{2\pi P_b} \int\int\int\int d\Omega d\Omega'd\Omega''d\Omega'''\\\nonumber
&\times&\left|J(\Omega,\Omega')J(\Omega'',\Omega''')\right|\\\nonumber
&\times&e^{\frac{iMD_{\mathrm{f}}}{2}[(\frac{\Omega}{M}-\Omega''')^{2}-(\frac{\Omega''}{M}-\Omega')^{2}]}\\\nonumber
&\times& e^{i(\Omega-\Omega'')\delta t-i(\Omega'-\Omega''')(\delta\tau-\delta t)}.
\end{eqnarray}
is the conditional probability of destructive interference of two photons under the condition that they are generated. By destructive interference we mean, in the spirit of Ref. \cite{Hong87}, the case where both photons go to the same beam-splitter output. The expression (\ref{pint}) looks complex, but with a simple argument, we can prove that it is real. Indeed, if we interchange the variables $\Omega$ and $\Omega'$ with $\Omega''$ and $\Omega'''$ respectively, then the expression becomes complex conjugate. At the same time, the change of variables should not change the integral; therefore we conclude that $p_\mathrm{int}(\delta\tau)$ is real. At sufficiently high values of $|\delta\tau|$ and $|\delta\tau-\delta t|$, the function under the integrals in Eq. (\ref{pint}) oscillates very fast with the frequency and its integral is close to zero. Thus, at a high enough delay, the probability of interference tends to zero and the total number of coincidences $TR_c(\pm\infty)$ tends to $P_b/2$, as expected since each photon can be reflected or transmitted at the beam splitter with a probability $\frac12$, giving four possible combinations, two of which result in a coincidence.

The quadruple integral (\ref{pint}) can be made Gaussian with the help of the Gaussian approximation for the phase-matching function (\ref{approxPhi}), already employed in Sec. \ref{sec:PDC} to calculate the spectral and temporal shapes of the ordinary and extraordinary waves. In this approximation, the integration can be taken analytically by the multidimensional Gaussian integration formula
\begin{equation}\label{multiGauss}
    \int_{-\infty}^{\infty} e^{-\mathbf{u}^T\Lambda \mathbf{u}/2+i\mathbf{v}^T\mathbf{u}} d^n u=\left[\frac{(2\pi)^n}{\det\Lambda}\right]^{\frac12} e^{-\mathbf{v}^T\Lambda^{-1} \mathbf{v}/2},
\end{equation}
where $\mathbf{u}$ and $\mathbf{v}$ are two $n$-dimensional column vectors, while $\Lambda$ is an $n\times n$ symmetric matrix.

To simplify the notation, we introduce dimensionless delays, relating them to the pump pulse duration $\tau_p$,
\begin{equation} \label{TOE}
  T_{o,e}=\frac{\sqrt{2}\Omega_p}{\sigma_s}\tau_{o,e}
  = \frac{2\sqrt{\ln2}}{\sigma_s}\frac{\tau_{o,e}}{\tau_p} \approx 1.034 \frac{\tau_{o,e}}{\tau_p}
\end{equation}
and a dimensionless focal GDD $D=2\Omega_p^2D_\mathrm{f}$.

To transform Eq. (\ref{pint}) with the phase-matching function approximated by Eq. (\ref{approxPhi}) into the matrix form of Eq. (\ref{multiGauss}), we identify $n=4$, $\mathbf{u}=(\Omega,\Omega',\Omega'',\Omega''')^T$, $\mathbf{v}=(\delta t, \delta t-\delta\tau, -\delta t, \delta\tau-\delta t)^T$, and
\begin{widetext}
\begin{equation}
    \Lambda=\frac1{2\Omega_p^2}
    \begin{pmatrix}
    1+T_o^2-iD/M & 1+T_oT_e & 0 & iD\\
    1+T_oT_e & 1+T_e^2+iDM & -iD & 0\\
    0 & -iD & 1+T_o^2+iD/M & 1+T_oT_e\\
    iD & 0 & 1+T_oT_e & 1+T_e^2-iDM\\
    \end{pmatrix}.
\end{equation}
\end{widetext}

We notice that the third and fourth elements of the vector $\mathbf{v}$ are equal to the first and second ones taken with the opposite sign. This allows us to simplify the quadratic form on the right-hand side of Eq. (\ref{multiGauss}) significantly. Let us denote the inverse of $\Lambda$ by $\Pi$ and write it in a block form
\begin{equation}
    \Pi =
    \begin{pmatrix}
    \Pi_{11} & \Pi_{12}\\
    \Pi_{21} & \Pi_{22}\\
    \end{pmatrix},
\end{equation}
where $\Pi_{ij}$ are some $2\times2$ matrices. Now we introduce a two-dimensional column vector $\mathbf{w}=(\delta t, \delta t-\delta\tau)^T$ and write
\begin{equation}
   \mathbf{v}^T\Lambda^{-1} \mathbf{v} = \mathbf{w}^T\Gamma \mathbf{w},
\end{equation}
where $\Gamma$ is a $2\times2$ matrix
\begin{equation}
  \Gamma =  \Pi_{11}- \Pi_{12}- \Pi_{21}+ \Pi_{22}.
\end{equation}

With the help of {\sc Mathematica 12}, we find
\begin{equation}\label{Gamma}
  \Gamma = \nu
  \begin{pmatrix}
    1+T_e^2+M^2F_+ & -1-T_eT_o-MF_+\\
    -1-T_eT_o-MF_+ & 1+T_o^2+F_+\\
    \end{pmatrix},
\end{equation}
where
\begin{eqnarray}
\nu &=& \frac{(T_e-T_o)^2}{4\Omega_p^6\det\Lambda},\\
\det\Lambda &=& \frac{(T_e-T_o)^4}{16\Omega_p^8}\left(1+M^2F_+F_-/D^2\right),
\end{eqnarray}
and
\begin{equation}
    F_\pm = \frac{D^2\left[(1\pm M)^2+(T_e\pm MT_o)^2\right]}{M^2(T_e-T_o)^2}.
\end{equation}

Thus, we finally obtain
\begin{equation}\label{J1fin}
   p_\mathrm{int}(\delta\tau) = \frac{\left|T_e-T_o\right| D_{\mathrm{f}}}{\Omega_p^2\sqrt{\det\Lambda}}
   e^{- \mathbf{w}^T\Gamma \mathbf{w}/2}.
\end{equation}

Let us show that both eigenvalues of $\Gamma$ are positive for any values of the parameters $T_e$, $T_o$, $D$, and $M$. First, we find
\begin{equation}
\det\Gamma = \nu^2 (T_e-T_o)^2\left(1+M^2F_+F_-/D^2\right),
\end{equation}
which is obviously positive. Then, we write the characteristic polynomial in the form
\begin{equation}
f(\lambda) = \lambda^2 - (\Gamma_{11}+\Gamma_{22})\lambda +\det\Gamma,
\end{equation}
where $\Gamma_{ij}$ are the elements of the matrix $\Gamma$. This function represents a parabola with a vertex at $\lambda_0=(\Gamma_{11}+\Gamma_{22})/2$, which is positive, where  $f(\lambda_0)=-(\Gamma_{11}+\Gamma_{22})^2/4-\Gamma_{12}^2$, which is negative. This means that the parabola intersects the $f=0$ axis at two points, one of which is positive. The second eigenvalue is positive too since the determinant is given by the product of eigenvalues. As a consequence, at high delay $\delta\tau$, $p_\mathrm{int}(\delta\tau)$ tends to zero, as anticipated above.

The average coincidence rate reaches its minimum at some delay $\delta\tau_\mathrm{min}$. This delay corresponds to the minimal value of the quadratic form $\mathbf{w}^T\Gamma \mathbf{w}$ and can be found by taking its derivative with respect to $\delta\tau$ and setting it to zero:
\begin{equation}\label{min}
   \delta\tau_\mathrm{min} = \left(1+
   \frac{\Gamma_{12}}{\Gamma_{22}}\right)\delta t.
\end{equation}

The visibility of the Hong-Ou-Mandel interference picture is defined as the ratio of the dip to the coincidence rate at a high delay:
\begin{equation}\label{Vis}
    V = \frac{R_c(\infty)-R_c(\delta\tau_\mathrm{min})}{R_c(\infty)}
    = p_\mathrm{int}(\delta\tau_\mathrm{min}),
\end{equation}
where we have used the expression for the average coincidence rate (\ref{Rc3}) and the fact that $p_\mathrm{int}(\infty)=0$, established above. Below we analyze the visibility for the cases of perfect and imperfect synchronization.

\subsection{Perfect synchronization}

At perfect synchronization of the PDC pump pulse and the time axis of the time lens, $\delta t=0$ and therefore, from Eq. (\ref{min}), $\delta\tau_\mathrm{min}=0$. The interference probability is
\begin{equation}\label{pintPS}
    p_\mathrm{int}^\mathrm{PS}(\delta\tau) = \frac{2D} {\left|T_e-T_o\right|\sqrt{1+M^2F_+F_-/D^2}} e^{-\Gamma_{22}\delta\tau^2/2},
\end{equation}
where
\begin{equation}
\Gamma_{22} = 4\sigma_\mathrm{cw}^2\frac{1+T_o^2+F_+}{1+M^2F_+F_-/D^2}.
\end{equation}

With varying magnification, the visibility $p_\mathrm{int}^\mathrm{PS}(0)$ reaches its maximum  at the minimal value of the function $g(M)=M^2F_+F_-$. Equalizing its derivative to zero, we find the optimal magnification
\begin{equation}\label{Mopt}
    M_\mathrm{opt}=\pm\sqrt{\frac{1+T_e^2}{1+T_o^2}} = \pm\frac{\Delta t_e}{\Delta t_o},
\end{equation}
where $\Delta t_o$ and $\Delta t_e$ are the standard deviations of the temporal shapes of the ordinary and extraordinary waves respectively, found in Sec.~\ref{sec:PDC}. The minimal value $g(M_\mathrm{opt})=4D^4/(T_e-T_o)^2$ and therefore the optimal visibility at perfect synchronization is
\begin{equation}
    V_\mathrm{opt}^\mathrm{PS} =  \frac{2D} {\sqrt{(T_e-T_o)^2+4D^2}}.
\end{equation}

From a physical point of view, the optimal magnification, given by Eq. (\ref{Mopt}), is completely clear and expected. It corresponds to the ratio of the temporal widths of the ordinary and extraordinary photons, i.e., at the optimal magnification, the duration of the stretched ordinary photon equals the duration of the extraordinary one. The minus sign in Eq. (\ref{Mopt}) corresponds to the inversion of the temporal waveform, which does not affect the visibility, since the photon has a symmetric temporal shape and we are considering a perfect synchronization, where the center of the photon coincides with the center of the field of view of the time lens.

\begin{figure}[ht]
\centering
\includegraphics[width=\linewidth]{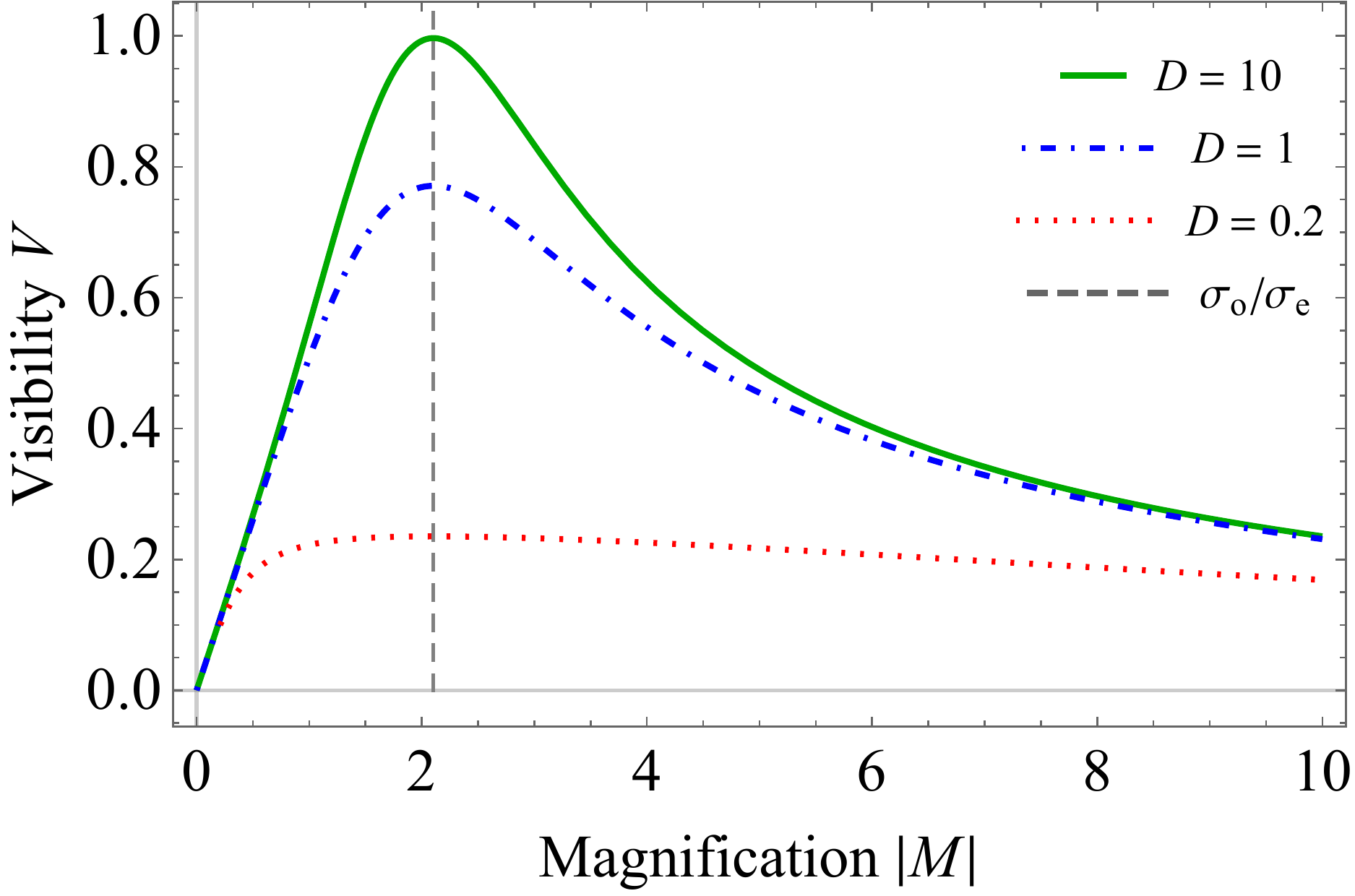}
\caption{Interferometric visibility plotted at perfect synchronization as a function of the magnification $|M|$ for different values of focal GDD. The source parameters are as in Sec. \ref{sec:shapes}. The maximal visibility is reached at the optimal magnification, corresponding to the ratio  $\Delta t_e/\Delta t_o=\sigma_o/\sigma_e$.  \label{fig:visibility2}}
\end{figure}

When the focal GDD of the time lens is high enough so that the relation $D^2\gg\left(T_e-T_o\right)^2/4$ or
\begin{equation}\label{HiD}
D_\mathrm{f}^2\gg\frac1{4\Omega_p^2\sigma_\mathrm{cw}^2}
\end{equation}
holds, the visibility approaches unity (see Fig.~\ref{fig:visibility2}).
At a lower $D$, the interference is not perfect because a time lens adds a residual chirp to the stretched ordinary pulse, which does not match perfectly the extraordinary one, notwithstanding having the same duration and the temporal intensity shape. This chirp tens to zero as $D_\mathrm{f}$ tends to infinity, as can be seen from Eq. (\ref{LensEq2}). For the examples considered in Sec.~\ref{sec:TI}, we obtain the values of $4D_\mathrm{f}^2\Omega_p^2\sigma_\mathrm{cw}^2$ equal to 0.55 and 650 for EOPM and FWM based lenses, respectively. We see that the FWM-based lens considered satisfies the condition (\ref{HiD}), while the EOPM-based lens considered does not reach the limit $D\to\infty$ and allows one to obtain only a limited visibility of $(1+1/0.55)^{-1/2}=0.6$.

\begin{figure}[ht]
\centering
\includegraphics[width=\linewidth]{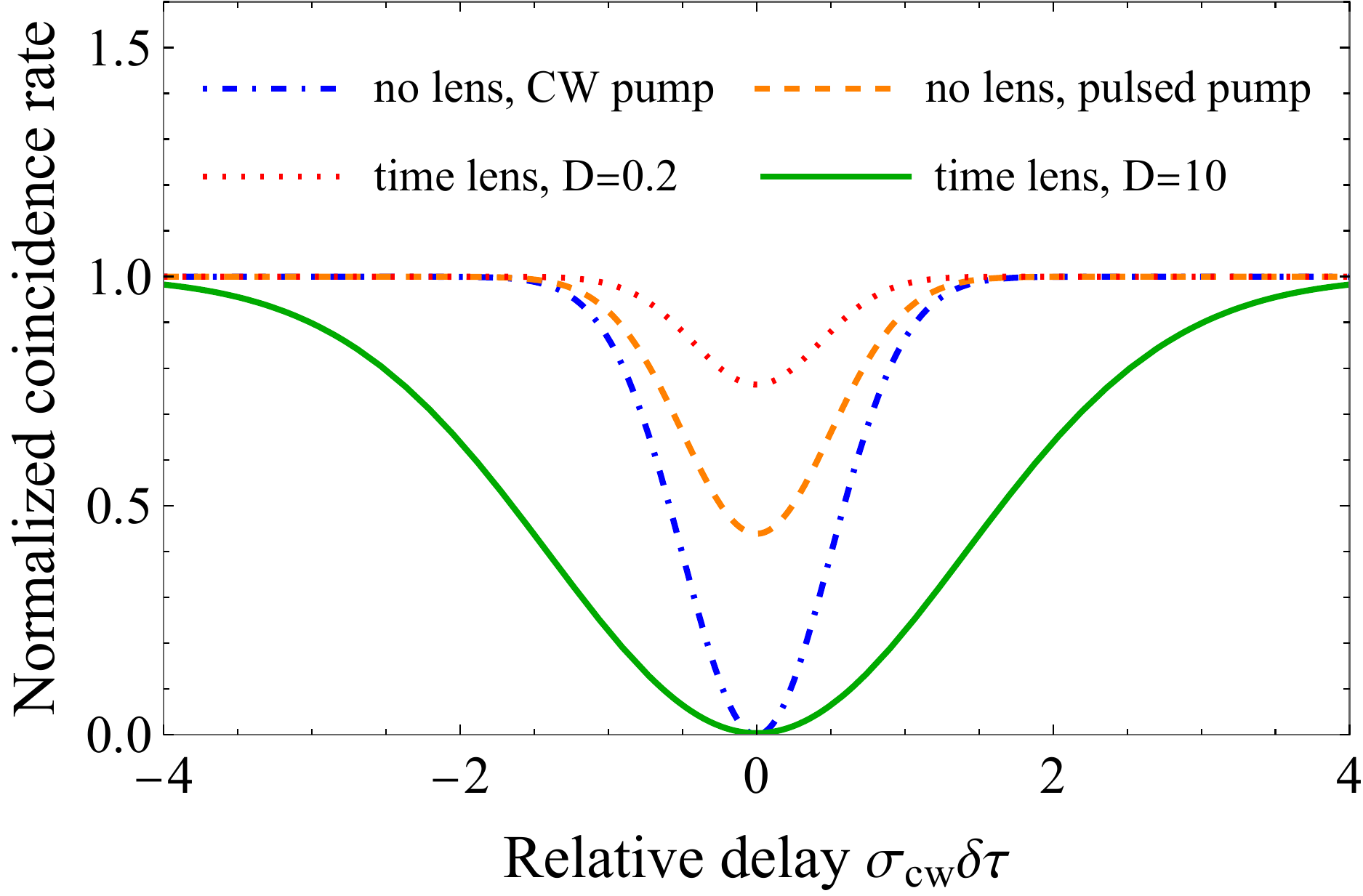}
\caption{Normalized coincidence rate plotted against the relative delay. The source parameters are as in Sec. \ref{sec:shapes}. The lines for a lensless case reproduce the well-known result: The pulsed pump reduces the visibility of Hong-Ou-Mandel interferometry \cite{Grice97,Keller97,Grice98}. A time lens with a low focal GDD is unable to restore the visibility, even at optimal magnification, because of the strong chirp, imposed on the magnified pulse. A time lens with a high focal GDD imposes a much weaker chirp and restores the visibility to 100\%. \label{fig:coincirate}}
\end{figure}

The limit of $M=1$ and $D_\mathrm{f}\to\infty$ corresponds to a lensless setup, where the imaging transformation (\ref{LensEq}) reduces to a translation in time. In this limit we find
\begin{equation}\label{lensless}
   p_\mathrm{int}^\mathrm{lensless}(\delta\tau) = \frac{2}{\sqrt{4+(T_e+T_o)^2}}
   e^{-2\sigma_\mathrm{cw}^2\delta\tau^2}.
\end{equation}
We see that for a lensless setup, the width of the HOM dip is independent of the pump spectral width, while the visibility is maximal in the CW limit ($\Omega_p\to0$), where it is equal to one, and degrades with the growing pump spectral width \cite{Grice97,Keller97}. In Fig.~\ref{fig:coincirate} we show the coincidence rate in the CW limit, its degraded version for a pulsed pump, and the interference picture restored by a time lens with a sufficiently high focal GDD.

We see in Fig.~\ref{fig:coincirate} that the dip of the restored interference picture is much wider than that of the case of the CW pump. This might appear paradoxical, because the CW limit corresponds to long pump pulses, e.g., nanosecond scale, which produce long subharmonic pulses, as discussed in Sec. \ref{sec:shapes}, while a short pump pulse produces subharmonic pulses of the same scale, picosecond in the considered example. It might be expected that longer pulses have a longer correlation time. However, that is not the case; the correlation time (standard deviation of the interference dip) in the lensless CW limit is equal to $\tau_\mathrm{cw}=0.5\sigma_\mathrm{cw}^{-1}$, as follows from Eq. (\ref{lensless}), and is unrelated to the pump pulse duration. For a pulsed pump and in the presence of a lens, the correlation time is $\tau_\mathrm{pulsed}=\Gamma_{22}^{-1/2}$, as follows from Eq. (\ref{pintPS}). In the example considered this time is $\tau_\mathrm{pulsed}=1.4\sigma_\mathrm{cw}^{-1}$ and is much longer than $\tau_\mathrm{cw}$.
A longer correlation time may be beneficial for some applications of indistinguishable photons, such as boson samplers \cite{Spring13,Shchesnovich20}, because it makes photons less sensitive to the emission time jitter.

\subsection{Imperfect synchronization}

\begin{figure}[ht]
\centering
\includegraphics[width=\linewidth]{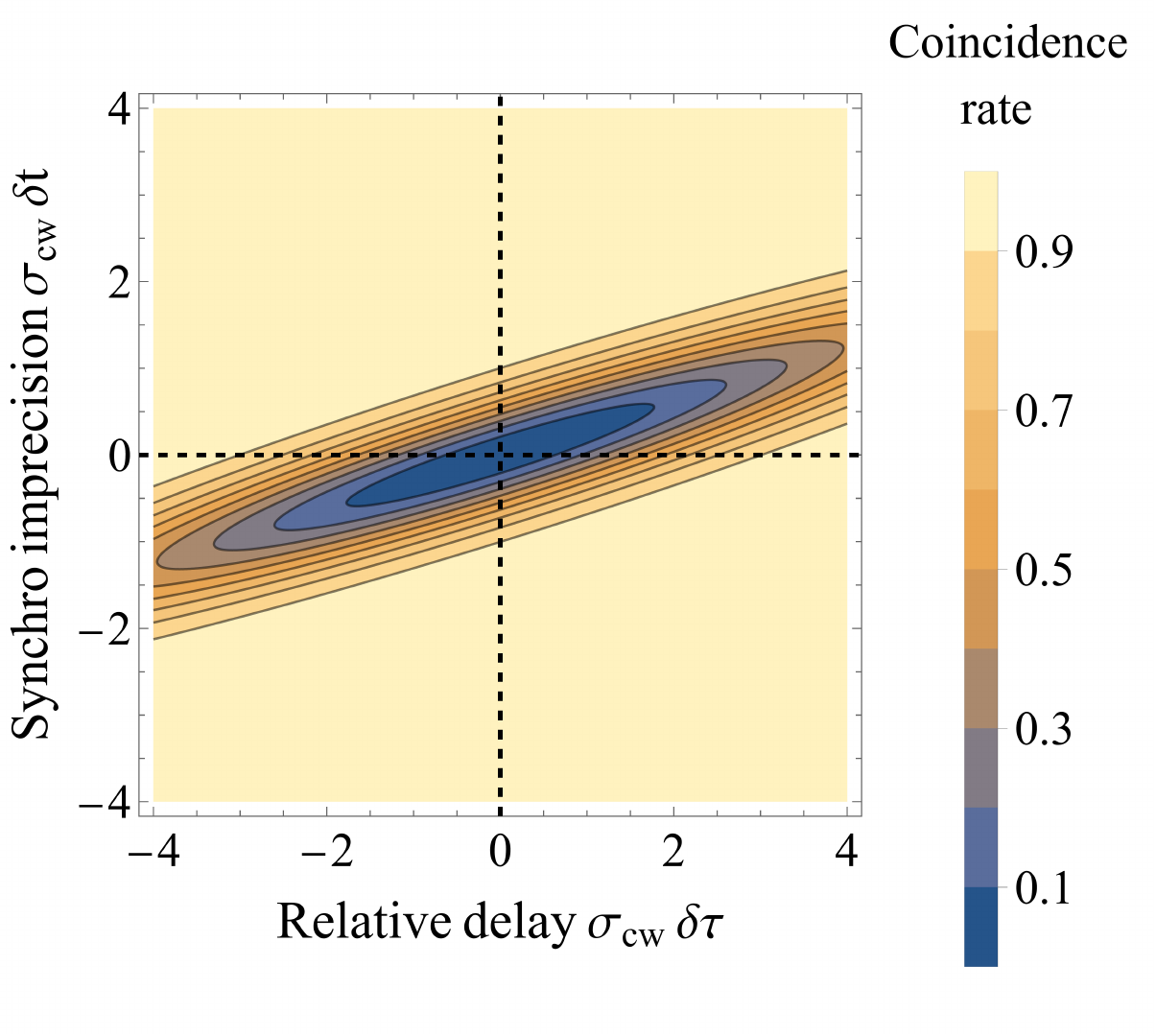}
\caption{Normalized coincidence rate plotted versus the synchronization imprecision $\delta t$ and the relative delay $\delta\tau$ measured in units of $\sigma^{-1}_{\text{cw}}$. The source parameters are as in Sec. \ref{sec:shapes}. The imaging system parameters are $D=10$ and $M=-2.1$. The dashed axes illustrate that the coincidence rate is minimal for $\delta t=0$ and $\delta\tau=0$.   \label{fig:coincirateDt}}
\end{figure}

In the case of imperfect synchronization, $\delta t \ne 0$ and the position of the dip of the average coincidence rate depends on $\delta t$ (see Fig.~\ref{fig:coincirateDt}). In addition,  this dip is less pronounced.

Substituting Eq. (\ref{min}) into Eq. (\ref{J1fin}) and the result into  Eq. (\ref{Vis}), we obtain the visibility
\begin{equation}
    V = \frac{2De^{-\gamma\delta t^2}}{\left|T_e-T_o\right|\sqrt{1+M^2F_+F_-/D^2}},
\end{equation}
where
\begin{equation}
    \gamma = \frac{2\Omega_p^2}{1+T_o^2+F_+}.
\end{equation}

\begin{figure}[ht]
\centering
\includegraphics[width=\linewidth]{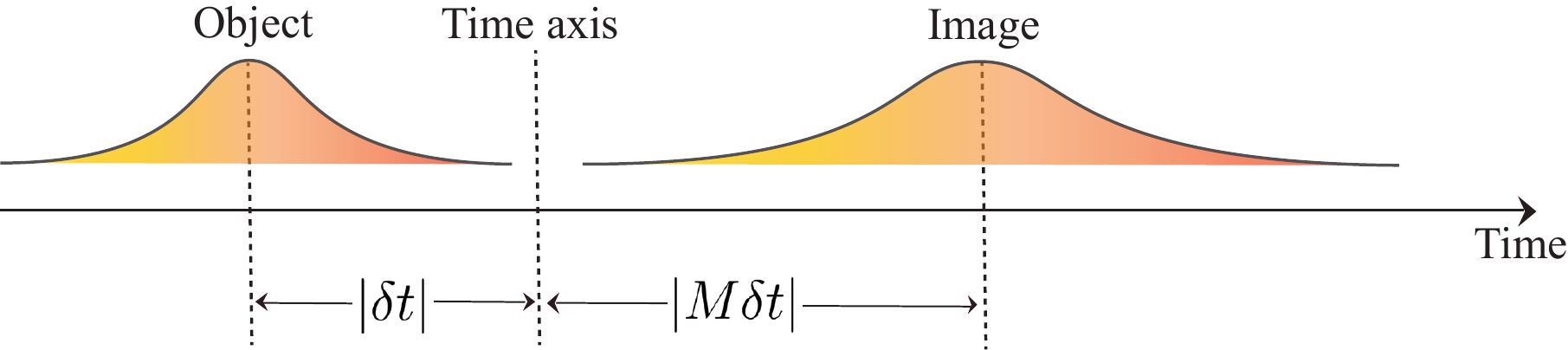}
\caption{Schematic representation of the effect of a time lens on an input waveform in the case of imperfect synchronization of the source and the time lens. The waveform center has a time imprecision $\delta t$ with the time axis of the lens. In the case of negative magnification, the output waveform is obtained by stretching $|M|$ times and inverting the input waveform. As a result, the magnified waveform experiences a time shift $(1+|M|)|\delta t|$. \label{fig:TimeEffect}}
\end{figure}

At high enough $D$, $\gamma$ tends to 0 and the visibility approaches that of perfect synchronization, analyzed in the preceding section. The delay corresponding to the dip (\ref{min}) takes a simple form in the limit $D\to\infty$:
\begin{equation}\label{impshift}
    \delta\tau_\mathrm{min} \to (1-M)\delta t.
\end{equation}
This result has a simple explanation. When the center of the single-photon pulse is delayed by $|\delta t|$ with respect to the time axis of the time lens with a magnification $M<0$, the image pulse is shifted at the output by $|M\delta t|$ in the opposite direction (Fig.~\ref{fig:TimeEffect}), which gives a total time shift of $(1+|M|)|\delta t|$. This shift is to be compensated by a time delay, specified in Eq.~(\ref{impshift}). For a positive $M$, the image lies at the same side of the time axis with the object, and the total time shift is $|(1-M)\delta t|$, again in agreement with Eq.~(\ref{impshift}).

The analysis of this section shows that, in general, the synchronization jitter reduces the visibility of the interference picture, but its effect can be greatly reduced by choosing a time lens with a sufficiently high focal GDD.

\subsection{Properties of photons after the lens \label{sec:after}}

To make more intuitive the conclusions made above, let us consider the spectral and temporal shapes of two photons at position $x_4$ just before the output beam splitter of the interferometer and compare them to those found in Sec. \ref{sec:shapes}. The extraordinary photon experiences just a delay, so its temporal and spectral widths are the same as at the crystal exit at position $x_1=L$. In contrast, the ordinary photon, which passes through the time lens, changes both its temporal and spectral shapes. The intensity of the ordinary wave at $x_4$ is $I_o^\mathrm{out}(t)=\langle\hat E_o^{(-)}(t,x_4)\hat E_o^{(+)}(t,x_4)\rangle$. Expressing the fields at $x_4$ through the fields at $x_1=L$ by the relations given in Sec. \ref{sec:SOI}, we find
\begin{equation}\label{Iout}
    I_o^\mathrm{out}(t) = \frac1{|M|} I_o\left(\frac{t-t_\mathrm{delay}}M\right),
\end{equation}
where the delay time is $t_\mathrm{delay}=t_{34}+t_\mathrm{out}+Mt_{12}-Mt_\mathrm{in}$. That is, the temporal width of the ordinary photon is increased $|M|$ times, as expected for a temporal imaging system with a magnification $M$.

To find the spectral shape of the ordinary photon we calculate the JSA at position $x_4$:
\begin{eqnarray}\label{JSAout}
    &&J_\mathrm{out}(\Omega,\Omega') =  \frac1{2\pi}\\\nonumber
    &&\times \int\langle \hat{E}^{(+)}_o(t,x_4)\hat{E}^{(+)}_e(t',x_4)\rangle e^{i(\omega_0+\Omega)t+i(\omega_0+\Omega')t'}dtdt'.
\end{eqnarray}
Expressing the fields at $x_4$ through the fields at $x_1=L$, as above, we obtain
\begin{equation}\label{JSAout2}
    J_\mathrm{out}(\Omega,\Omega') =  e^{i(k_e'L+t_z)\Omega'}\int G_o(\Omega,\bar\Omega)J(\bar\Omega,\Omega')\frac{d\bar\Omega}{2\pi},
\end{equation}
where
\begin{equation}\label{Go}
    G_o(\Omega,\bar\Omega) = \sqrt{2\pi iD_\mathrm{f}}e^{-\frac{i}2MD_\mathrm{f}\left(\Omega-\frac{\bar\Omega}M\right)^2+i\Omega\tau_2+i\bar\Omega\tau_1+i\phi'}
\end{equation}
is the transfer function for the ordinary photon in the frequency domain. Here $\tau_2=t_\mathrm{out}+t_{34}$, $\tau_1=k_o'L+t_{12}-t_\mathrm{in}$, and $\phi' = \phi + k_e^0L+\omega_0t_z$. The transfer function satisfies the unitarity condition
\begin{equation}\label{GG}
    \int G_o^*(\Omega,\Omega_1) G_o(\Omega,\Omega_2)\frac{d\Omega}{2\pi} = 2\pi\delta(\Omega_1-\Omega_2).
\end{equation}
As a consequence, the spectral and temporal shapes of the extraordinary photon, calculated from equations similar to Eqs. (\ref{So}) and (\ref{Io}), remain unchanged except for a delay, as anticipated above.

The spectral width of the ordinary photon can be easily found in the limit $D_\mathrm{f}\to\infty$, where $G_o(\Omega,\bar\Omega)\to \frac{2\pi}{\sqrt{M}}\delta(\Omega-\bar\Omega/M)$ and
\begin{equation}\label{Joutlim}
    J_\mathrm{out}(\Omega,\Omega') \to  \sqrt{M}J(M\Omega,\Omega'),
\end{equation}
which means that the spectral width of the ordinary photon decreases $|M|$ times.

To find the spectrum of the ordinary photon in the general case of finite $D_\mathrm{f}$, we perform a Gaussian modeling of the phase-matching function, as in Sec. \ref{sec:shapes}. Substituting Eqs. (\ref{alpha}), (\ref{J}), and (\ref{approxPhi}) into Eq. (\ref{JSAout2}), we obtain
\begin{equation}\label{Joutfin}
    J_\mathrm{out}(\Omega,\Omega') = J_0 e^{-\frac{1}{4\Sigma^2}\left(B_{11}\Omega^2+2B_{12}\Omega\Omega' +B_{22}\Omega^{'2}\right)+i\psi(\Omega,\Omega')},
\end{equation}
where $B_{11}=D^2(1+T_o^2)$, $B_{12}=D^2(1+T_oT_e)/M$, and
$B_{22}=D^2(1+T_e^2)/M^2+(T_e-T_o)^2(1+T_o^2)$, while the values of $J_0$, $\Sigma$ and $\psi(\Omega,\Omega')$ can be found in Appendix \ref{sec:appendixc}. We see that the spectral correlations (entanglement) between the photons are still present after the time lens, as indicated by the nonvanishing coefficient $B_{12}$.

The output JSA is evidently nonsymmetric at the optimal magnification, given by Eq.~(\ref{Mopt}). Thus, the spectral widths of the photons are different at the optimal magnification for finite $D_\mathrm{f}$. In the case $|D|>|T_o-T_e|$, the modulus of the output JSA can be made symmetric by choosing the magnification
\begin{equation}\label{Msym}
    M_\mathrm{sym} = \pm\frac{M_\mathrm{opt}}{\sqrt{1-(T_e-T_o)^2/D^2}},
\end{equation}
which is, however, suboptimal.

Substituting Eq. (\ref{Joutfin}) into Eq. (\ref{So}), we obtain the spectrum of the ordinary photon after the time lens in a form of a Gaussian distribution with the dispersion
\begin{equation}\label{sigmaoout}
    \sigma_{o,\mathrm{out}}^2 = \frac{\Omega_p^2}{D^2}\frac{\left[D^2/M^2+(1+T_o^2)^2\right](1+T_o^2)\eta}{(1+T_o^2)\eta-(1+T_oT_e)^2/M^2},
\end{equation}
where $\eta=M_\mathrm{opt}^2/M^2+(T_e-T_o)^2/D^2$. At the optimal magnification and in the limit of high $D$, $\eta\to1$ and $\sigma_{o,\mathrm{out}}\to\sigma_o/|M|=\sigma_e$, as expected.

Thus, a single-lens temporal imaging system with a magnification $M$ realizes a scaling of the field with the factor $|M|$ in the temporal domain: A pulse of duration $\Delta t$ is stretched to the duration $|M|\Delta t$. However, in the spectral domain, the scaling factor is not given by $|M|$: A pulse of spectral width $\sigma$ is transformed into a pulse with a spectral width higher than $\sigma/|M|$ because of an additional chirp imposed by the temporal imaging system. It is interesting that the highest visibility is attained at a magnification corresponding to equal durations of two photons, not to their equal spectral widths.

\section{Conclusion}\label{sec:D}

We have considered an application of a time lens for the light produced in a frequency-degenerate type-II SPDC pumped by a pulsed broadband source. In this type of SPDC, the signal and the idler photons have different spectral and temporal properties. Therefore, they lose their indistinguishability, and this effect deteriorates the visibility of the Hong-Ou-Mandel interference. We have demonstrated that by inserting a time lens in one of the arms of the interferometer and choosing appropriately its magnification factor one can achieve 100\% visibility in the Hong-Ou-Mandel interference and, thus, restore perfect indistinguishability of the signal and idler photons.

In our theoretical model, we have taken into account the fact that for a pulsed source of the SPDC light one has  to properly synchronize in time the source and the time lens in order to achieve the optimum destructive interference of the signal and the idler waves. We have studied the effect of desynchronization on the visibility of second-order interference. Our results show that for large values of the focal GDD of the lens the effect of the desynchronization can be neglected and one can still have perfect visibility of the interference, however, for a nonzero time delay in the interferometric scheme.

In our calculations, we have assumed a Gaussian shape of the pump pulses for the SPDC source and also used a Gaussian approximation for the phase-mismatch  function of the nonlinear crystal. These two conditions have allowed us to obtain several important analytical results, for example, for the spectra of the ordinary and the extraordinary waves and also for the temporal shapes of the emitted pulses. We have also obtained an analytical formula for the optimum value of the magnification factor of the time lens, which provides the unit visibility of the Hong-Ou-Mandel interference. We have given a simple physical interpretation of this result in terms of the temporal durations of the ordinary and the extraordinary pulses.

In order to assess the feasibility of our theoretical proposal in view of the possible experimental realizations, we have provided quantitative estimations for two possible implementations of the time lens: based on EOPM, and on FWM. Taking as an example a BBO crystal for the SPDC light source, we have shown that for both types of time lens our theoretical proposal is experimentally feasible, though for an EOPM-based time lens, the range of possible input bandwidths is much shorter than for a FWM-based time lens.

The single-lens scheme considered in this paper does not allow one to completely eliminate the residual phase chirp in the temporal image. Our calculations show, however, that this residual chirp becomes negligible for large values of the focal GDD of the lens, easily reachable for FWM, but not so easily accessible for a more practical EOPM. Extension of our scheme to a time telescope comprising two time lenses, convergent and divergent, is left for future work. Such a telescope allows one to create a temporal image without phase chirp. It should therefore be effective in reestablishing the indistinguishability of the signal and the idler photons for arbitrary values of the focal GDD of two lenses constituting the telescope.

\begin{acknowledgments}
This work was supported by the network QuantERA of the European Union’s Horizon 2020 research and innovation program under the project Quantum information and communication with high-dimensional encoding , the French part of which is funded by Agence Nationale de la Recherche via Grant No. ANR-19-QUAN-0001.
\end{acknowledgments}

\appendix
\section{Photon spectra and intensities \label{sec:appendixa}}

Substituting Eqs. (\ref{alpha}), (\ref{approxPhi}) and (\ref{J}) into Eq. (\ref{So}), we obtain

\begin{align} \label{So2}
    S_o (\Omega) &=\pi\left(\frac{E_0 \kappa L}{\Omega_p}\right)^{2}\int e^{-\frac{(\Omega+\Omega')^2}{2\Omega_p^2} -\frac{(\tau_o\Omega+\tau_e\Omega')^2}{\sigma_s^2}} d\Omega'.
\end{align}

This expression is a one-dimensional Gaussian integral and can be evaluated by using the formula
\begin{equation}\label{GaussFormula}
    \int_{-\infty}^{\infty}e^{- a x^2 + b x}\,dx=\sqrt{\frac{\pi}{a}}\,e^{b^2/4a}.
\end{equation}
In this way, we obtain Eq. (\ref{Smu}).

Similarly, substituting Eqs. (\ref{alpha}), (\ref{approxPhi}) and (\ref{J}) into Eq. (\ref{Io}), we obtain
\begin{eqnarray}\label{Ioapp}
I_o(t) &=& \left(\frac{\kappa LE_0}{\Omega_p}\right)^2 \frac1{4\pi}\int e^{i(\Omega-\Omega')(t-t_0-\tau_o-k_o'L)}\\\nonumber
&\times& \exp\left[-\frac{(\tau_o\Omega+\tau_e\Omega'')^2+(\tau_o\Omega'+\tau_e\Omega'')^2}{2\sigma_s^2}\right]\\\nonumber
&\times& \exp\left[-\frac{(\Omega+\Omega'')^2+(\Omega'+\Omega'')^2}{4\Omega_p^2}\right]d\Omega d\Omega' d\Omega''.
\end{eqnarray}
Making a change of variables $\Omega_+=\Omega+\Omega'$ and $\Omega_-=\frac12(\Omega-\Omega')$, we rewrite Eq. (\ref{Ioapp}) as
\begin{equation}\label{Ioapp2}
I_o(t) = \tilde{I}\int e^{2i\Omega_-(t-t_0-\tau_o-k_o'L) -\Omega_-^2/2\Sigma_o^2}d\Omega_-,
\end{equation}
where $\Sigma_o^{-2}=\Omega_p^{-2}+2\tau_o^2\sigma_s^{-2}$
%
%
and
\begin{eqnarray}\label{Itilde}
\tilde{I} &=& \left(\frac{\kappa LE_0}{\Omega_p}\right)^2 \frac1{4\pi} \int \exp\left[-\frac{(\frac12\Omega_++\Omega'')^2}{2\Omega_p^2}\right]\\\nonumber
&\times& \exp\left[-\frac{(\frac12\Omega_+\tau_o+\Omega''\tau_e)^2}{\sigma_s^2}\right]d\Omega_+d\Omega''.
\end{eqnarray}
The integral in Eq. (\ref{Ioapp2}) is evaluated by applying Eq. (\ref{GaussFormula}) and that in Eq. (\ref{Itilde}) -- by applying this formula twice. As a result, we obtain Eq. (\ref{Imu}).

\section{Fraunhofer limit for dispersion of a chirped pulse \label{sec:appendixb}}

In most publications on a time lens, a Fourier-limited pulse is considered at the input to the temporal imaging system. However, a single-photon pulse produced by a PDC source in a sufficiently long crystal may be not Fourier-limited, having a nonnegligible frequency chirp. Here we analyze the dispersive broadening of a chirped pulse and find the conditions for the Fraunhofer limit of dispersion.

We consider a classical field with the positive-frequency part $E^{(+)}(t) = Y_0(t)e^{-i\omega_0t}$, where $\omega_0$ is the carrier frequency and $Y_0(t)$ is the envelope amplitude, which we assume to have a Gaussian distribution with the (intensity) standard deviation $\Delta t_0$:
\begin{equation}
    Y_0(t) = E_0e^{-t^2/4\Delta t_0^2},
\end{equation}
where $E_0$ is the peak amplitude. In the frequency domain, we have
\begin{equation}
    \tilde Y_0(\Omega) = \int Y_0(t) e^{i\Omega t}dt = 2\sqrt{\pi}\Delta t_0E_0e^{-\Delta t_0^2\Omega^2},
\end{equation}
so the standard deviation of the intensity spectrum $S_0(\Omega) = |\tilde Y_0(\Omega)|^2$ is $\sigma_0=1/2\Delta t_0$. For the FWHM temporal and spectral widths $T_0^\mathrm{F}$ and $\Omega_0^\mathrm{F}$, respectively, we find the time-bandwidth product $T_0^\mathrm{F}\Omega_0^\mathrm{F} = 4\ln2\approx 2\pi\times0.44$, as expected for a Fourier-limited Gaussian pulse.

This pulse passes through a dispersive medium of length $L_1$ with the dispersion law $k_1(\Omega) = k_1^0+k_1'\Omega+k_1''\Omega^2/2$, which we limit to terms up to the quadratic one in the frequency detuning $\Omega=\omega-\omega_0$. The field at the output is simply $\tilde Y_1(\Omega) = \tilde Y_0(\Omega)e^{ik_1(\Omega)L_1}$, which gives in the time domain
\begin{equation}
    Y_1(t) = \int \tilde Y_1(\Omega) e^{-i\Omega t}\frac{d\Omega}{2\pi}
    = E_1e^{-\tau^2/4\Delta t_1^2 - i\tau^2/2C_1},
\end{equation}
where $E_1=E_0e^{ik_1^0L_1}/\sqrt{1-iD_1/2\Delta t_0^2}$ is the new peak amplitude, $D_1=k_1''L_1$
is the GDD acquired in the medium, $\tau=t-k_1'L_1$ is the time delayed by the group delay in the medium,
\begin{equation}\label{C1}
C_1 = D_1+4 \Delta t_0^4/D_1
\end{equation}
is the chirp coefficient of the output pulse, and
\begin{equation}\label{Deltat1}
    \Delta t_1 = \sqrt{\Delta t_0^2+D_1^2/4\Delta t_0^2}
\end{equation}
is the (intensity) standard deviation of the temporal distribution of the outgoing pulse. The spectral distribution is unchanged $|\tilde Y_1(\Omega)|^2 = |\tilde Y_0(\Omega)|^2$ and the spectral standard deviation is $\sigma_1=\sigma_0$.

Had we considered the Fraunhofer limit for this dispersion, we would have obtained the condition $D_1^2\gg4\Delta t_0^4$ and the relation $\Delta t_1 = |D_1|/2\Delta t_0$ \cite{Copmany11,Kolner94,Salem:13,Karpinski17}. However, in our case, we are looking for the Fraunhofer limit for the second dispersive medium of length $L_2$ with the dispersion law $k_2(\Omega) = k_2^0+k_2'\Omega+k_2''\Omega^2/2$ on the path of the same pulse. Similarly to the first medium, we have for the pulse at the output of the second medium:
$\tilde Y_2(\Omega) = \tilde Y_1(\Omega)e^{ik_2(\Omega)L_2} =\tilde Y_0(\Omega)e^{i[k_1(\Omega)L_1+k_2(\Omega)L_2]}$, which gives in the time domain
\begin{equation}
    Y_2(t) = E_2e^{-\tau^2/4\Delta t_2^2 - i\tau^2/2C_2},
\end{equation}
where $E_2=E_0e^{i(k_1^0L_1+k_2^0L_2)}/\sqrt{1-i(D_1+D_2)/2\Delta t_0^2}$ is the new peak amplitude, $D_2=k_2''L_2$
is the GDD aquired in the second medium, $\tau=t-k_1'L_1-k_2'L_2$ is the time delayed by the total group velocity delay in both media, $C_2 = D_1+D_2+4 \Delta t_0^4/(D_1+D_2)$ is the chirp coefficient of the output pulse, and
\begin{equation}\label{Deltat2}
    \Delta t_2 = \sqrt{\Delta t_0^2+(D_1+D_2)^2/4\Delta t_0^2}
\end{equation}
is the (intensity) standard deviation of the temporal distribution of the outgoing pulse after the second medium.

The Fraunhofer limit for the dispersion in the second medium occurs when the second term under the square root in Eq. (\ref{Deltat2}) is much greater than the first one, meaning $(D_1+D_2)^2\gg4\Delta t_0^4$, in which case we can rewrite Eq. (\ref{Deltat2}) as $\Delta t_2 = |D_1+D_2|/2\Delta t_0$. We can rewrite these two relations through the characteristics of the pulse entering the second medium only, if we express $(\Delta t_0,D_1)$ via $(\Delta t_1,C_1)$ by inverting the system of equations (\ref{C1}) and (\ref{Deltat1}). For this end, we multiply Eq. (\ref{C1}) by $D_1$ and the square of Eq. (\ref{Deltat1}) by $4\Delta t_0^2$ and obtain the same expression on the right-hand sides, i.e.,
$C_1D_1 = 4\Delta t_0^2 \Delta t_1^2$.
This expression allows us to exclude $D_1$ from Eq. (\ref{Deltat1}) and $\Delta t_0$ from Eq. (\ref{C1}), obtaining $\Delta t_0=1/2\sigma_0=1/2\sigma_1 =\Delta t_1(1+4\Delta t_1^4/C_1^2)^{-1/2}$ and $D_1=C_1(1+C_1^2/4\Delta t_1^4)^{-1}$. The two relations of the Fraunhofer dispersion limit are thus
\begin{eqnarray}\label{Fraunhofer}
    (D_1+D_2)^2 &\gg& 1/4\sigma_1^4 =4\Delta t_1^4(1+4\Delta t_1^4/C_1^2)^{-2}, \\\label{Deltat2bis}
    \Delta t_2 &=& |D_1+D_2|\sigma_1 \\\nonumber &=& |D_1+D_2|(1+4\Delta t_1^4/C_1^2)^{1/2}/2\Delta t_1.
\end{eqnarray}

We see that for a considerable $D_1$, the condition $D_2^2\gg4\Delta t_1^4$ may be not satisfied and $\Delta t_2 \ne D_2/2\Delta t_1$ in the Fraunhofer limit. Instead, Eqs. (\ref{Fraunhofer}) and (\ref{Deltat2bis}) should be used. In our setup, $D_1$ is the GDD acquired by the ordinary photon in the nonlinear crystal and $D_2$ is the GDD of the dispersive element before the time lens, and we define $D_\mathrm{in}=D_1+D_2$.

\section{Parameters of JSA after the lens \label{sec:appendixc}}
Substituting Eqs. (\ref{alpha}), (\ref{J}) and (\ref{approxPhi}) into Eq. (\ref{JSAout2}) and taking a Gaussian integral with the help of Eq. (\ref{GaussFormula}), we obtain
Eq. (\ref{Joutfin}) with $\Sigma^2=\Omega_p^2\left[(1+T_o^2)^2+D^2/M^2\right]$, $J_0=\kappa LE_0e^{i\phi'}\sqrt{2\pi iD_\mathrm{f}/(1+T_o^2+iD/M)}$ and
\begin{eqnarray}\label{psi}
    \psi(\Omega,\Omega') &=& \tau_2(\Omega+\Omega')\\\nonumber
    &-&\frac{D\left[M(1+T_o^2)\Omega+(1+T_oT_e)\Omega'\right]^2}{4M\Sigma^2}.
\end{eqnarray}
The phase of the JSA, $ \psi(\Omega,\Omega')$, becomes a symmetric function of its arguments in the limit $D\to\infty$, similar to the modulus of the JSA, analyzed in Sec. \ref{sec:after}.

\bibliography{Temporal-Imaging5}
\end{document}